\documentclass[aps,prl,reprint,superscriptaddress]{revtex4-1}

\usepackage[utf8]{inputenc}
\usepackage{graphicx}
\usepackage{dcolumn}
\usepackage{bm}
\usepackage{amssymb}
\usepackage{color}
\usepackage{ulem}
\usepackage{dsfont}

\newcommand{\cf}{{\it cf. }}

\newcommand{\vantri}{{V$_{2}$O$_{3}$ }}
\newcommand{\vandi}{{VO$_{2}$ }}
\newcommand{\celsius}{\,$^\circ$C\ }

\begin{document}


\title{Electrical Breakdown in a \vantri device at the Insulator to Metal Transition}


\author{S. Gu\'enon}
\email[]{stefan.guenon@physics.ucsd.edu}
\affiliation{Department of Physics and Center for Advanced Nanoscience, University of California-San Diego, La Jolla, California 92093, USA}
\author{S. Scharinger}
\affiliation{Physikalisches Institut and Center for Collective Quantum
  Phenomena in LISA$^+$, Universität Tübingen, Auf der Morgenstelle 14, D-72076 Tübingen, Germany}
\author{Siming Wang}
\affiliation{Department of Physics and Center for Advanced Nanoscience,
  University of California-San Diego, La Jolla, California 92093, USA}
\affiliation{Materials Science and  Engineering Program, La Jolla, California 92093, USA}
\author{J. G. Ram\'irez}
\affiliation{Department of Physics and Center for Advanced Nanoscience, University of California-San Diego, La Jolla, California 92093, USA}
\author{D. Koelle}
\affiliation{Physikalisches Institut and Center for Collective Quantum
  Phenomena in LISA$^+$, Universität Tübingen, Auf der Morgenstelle 14, D-72076 Tübingen, Germany}
 \author{R. Kleiner}
\affiliation{Physikalisches Institut and Center for Collective Quantum
  Phenomena in LISA$^+$, Universität Tübingen, Auf der Morgenstelle 14, D-72076 Tübingen, Germany}
\author{Ivan K. Schuller}
\affiliation{Department of Physics and Center for Advanced Nanoscience, University of California-San Diego, La Jolla, California 92093, USA}


\date{\today}

\begin{abstract}
We have measured the electrical properties of a \vantri thin film micro bridge
at the insulator metal transition (IMT). 
Discontinuous jumps to lower voltages in the current voltage characteristic
(IV) followed by an approximately constant voltage progression for
high currents indicate an electrical breakdown of the device. 
In addition, the IV curve shows hysteresis and a training effect, i.e. the
subsequent IV loops are different from the first IV loop after thermal cycling.  
Low temperature scanning electron microscopy (LTSEM) reveals that the
electrical breakdown over the whole device is caused by the formation of electro-thermal domains
(ETDs), i.e. the current and temperature redistribution in the device.
On the contrary, at the nanoscale, the electrical breakdown causes the IMT
of individual domains.
In a numerical model we considered these domains as a network of resistors and
we were able to reproduce the electro-thermal breakdown as well as the
hysteresis and the training effect in the IVs.
\end{abstract}

 \maketitle

\section{Main Paper}
Stoichiometric \vantri is a strongly correlated material that undergoes a
first order insulator to metal phase transition (IMT) from an
antiferromagnetic insulating to a paramagnetic metallic phase.
Because the IMT causes a change in the resistivity of several orders of
magnitude novel devices based on the IMT and their applications are actively
investigated \cite{Yang2011} and it is in focus of current research whether
there is a voltage driven IMT in strongly correlated materials \cite{Oka2003, Okamoto2008,
  Eckstein2010, Heidrich-Meisner2010, Liu2012}.
If this hypothesis holds, a micro bridge fabricated of a strongly correlated
material would experience a dielectric breakdown for high bias voltages. 
However, because of the large resistivity change, self-heating in such a
device can be strong enough to cause an electro-thermal breakdown, i.e. an 
current and temperature redistribution \cite{Berglund1969, Duchene1971}.
In order to investigated the electrical breakdown of a \vantri micro bridge, we used a  very unique low temperature scanning electron microscope (LTSEM) to image
the metallic and insulating phases and developed a numerical model  to
simulate the electrical device properties.
Here we show that the electrical breakdown over the whole device is due to
the formation of electro-thermal domains (ETDs), and that the distribution in
the IMT temperature as well as the hysteresis in the RT of domains at the nanoscale
are significantly influencing the current voltage characteristic (IV).\\ 
%

\begin{figure}
\includegraphics[width=\linewidth]{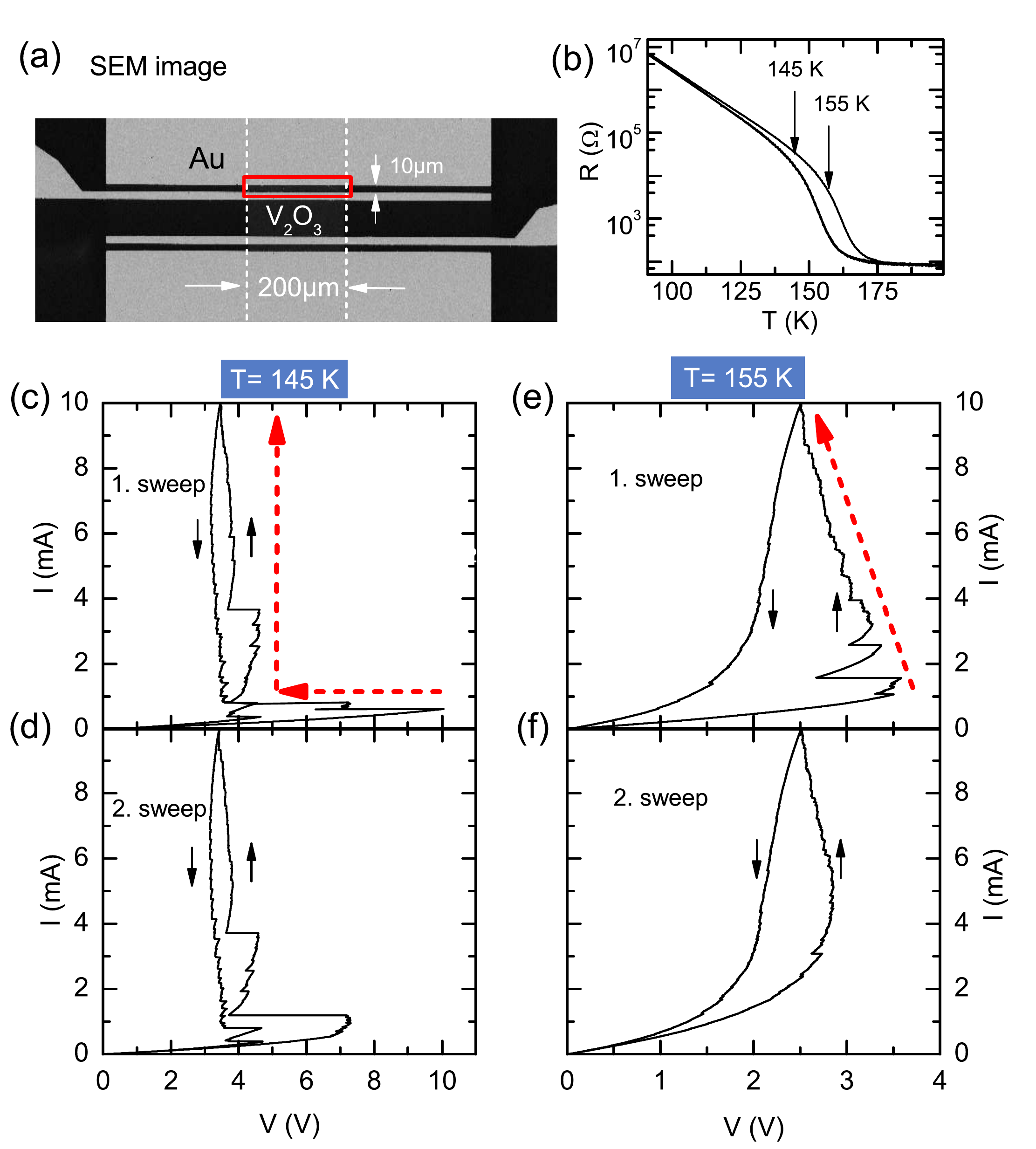}
\caption{\label{fig:iv}
(Color online)
Electrical breakdown. 
(a) SEM image of the device. 
The area under investigation is indicated by a red rectangle.  
A $200\,\mu$m wide \vantri stripe runs vertically indicated by two
white dashed lines. 
The two gold electrodes at the top form a $10\,\mu$m wide gap.
The current flows between those two electrodes vertically.
(b) Temperature dependence of the device resistance.
Current voltage characteristics (IVs) at two different base temperatures. The
black arrows indicate the sweep direction. The overall progression of the
electrical breakdown is indicated by red dashed arrows. 
(c) and (e) are IVs of the pristine device after cooling to 80 K. 
(d) and (f) are subsequent IVs to (c) and (e), respectively.}
\end{figure}

%
The \vantri film was grown by rf-sputtering on a r-cut sapphire substrate at a
temperature of approximately 750\celsius.
XRD revealed that the polycrystalline film grows textured under
these conditions.
Optical contact lithography was used for patterning.
A 200 $\mu\rm{m}$ wide \vantri bridge was etched using reactive ion
etching and gold electrodes were evaporated on top, refer to Fig. \ref{fig:iv} (a) for
an SEM image of the device.\\
The same experimental set up was used for LTSEM imaging and IV characterization.
The LTSEM is a conventional, state of the art SEM equipped with a LN$_2$-cryostat.
The sample is mounted in a vacuum on a cold plate and the sample temperature
can be varied from room temperature to approximately 80\,K.
Due to the large working distance required by the IR-shielding  of the
cryostat the smallest spot size of the electron beam is about $300\,$nm.
A Keithley 2400 current source and a conventional preamplifier with a National
Instruments measurement card were used for the two point electrical
measurements.\\
To acquire a LTSEM image, the sample is scanned with a blanked
electron beam and the electric response $\Delta V$ of a simultaneously applied
four probe or two probe measurement is mapped in a so called voltage image of
the sample \cite{Clem1980, Gross94}.\\ 
%
We have measured the current-voltage characteristics of a 200\,$\mu$m wide
and 10\,$\mu$m long \vantri bridge (Fig. \ref{fig:iv}a) for the base temperatures
$T_b=145\,$K and $T_b=155\,$K, respectively.
Due to the IMT there is a resistance change by more than
three orders of magnitude in a 20\,K temperature interval (Fig. \ref{fig:iv}b) 
and a pronounced hysteresis, i.e. the heating curve is shifted to higher
temperatures by approximately 5\,K.\\
To ensure that the \vantri film was completely in the insulating state the
device temperature was reduced to approximately 80\,K and then heated to the intended base
temperature before electrical measurements were started.
In the following, we refer to this procedure as ``thermal cycling''.\\
Fig. 1c shows an IV of the pristine device (after thermal cycling) acquired at
a base temperature T$_b$=145\,K.
Starting at the origin of the graph, the IV shows an almost linear dependence with a slight upwards bent.
At a current value of 0.6\,mA there is a jump from 10\,V to 7\,V followed by several
small jumps and a big jump to 3.8\,V. 
The curve continues almost vertically with a positive slope and several
sawtooth like discontinuities of different sizes.
The down sweep curve is different from the up sweep, i.e. there is a
hysteresis.
Immediately after the first IV curve, without changing the base temperature, a
second IV was acquired (Fig. \ref{fig:iv}f).
Note, that the IV of the second sweep is very similar to the first one except
that the maximum voltage is significantly reduced.
When the device is thermally cycled, the old state (first sweep) is restored.
Hence, there is a training effect.\\
The curve acquired at the first sweep at 155\,K (Fig. \ref{fig:iv}e) is
different from the one at 145\,K.
Starting at the origin, the IV is almost linear. 
After 1\,mA and 3.5\,V it progresses with a negative slope and saw tooth like
discontinuities of different sizes.
The down sweep curve is rounded with some very small discontinuities and the
hysteresis is much more pronounced than at 145\,K.
In the second IV (Fig.\ref{fig:iv}f) the down sweep is similar to the down
sweep of the first IV, but the up sweep curve is round and there are only a
few small discontinuities.
A comparison between the IV of the first and second sweep reveals, that the
training effect at 155\,K leads to a qualitative change of the up sweep curve.
The consecutive IVs do not change significantly after the second sweep at both
base temperatures.
Moreover, we have observed that the details of the IV depend on the sweep
velocity.
If the current is swept faster, there are more and smaller jumps, but 
the overall shape of the IV stays the same.\\
At T$_b$=145\,K a big jump in the IV followed by a vertical progression
indicates an electrical breakdown of the device.
The diagonal progression with a negative slope and small jumps in the first IV
at T$_b$=155\,K implies that the electrical breakdown now evolves via stable 
intermediate states. 
From the hysteresis in the IVs and the training effect it can be inferred that
there is a memory effect in the \vantri device, i.e. the device properties
depend on its history.
Similar effects were reported in \vandi devices \cite{Kim2010}.
The slight influence of the sweep velocity on the IV might be caused by a
slowed down relaxation of a small fraction of the film due to the spread in
the IMT temperatures of individual domains \cite{Grygiel2008}.\\
\begin{figure}  
\includegraphics[width=\linewidth]{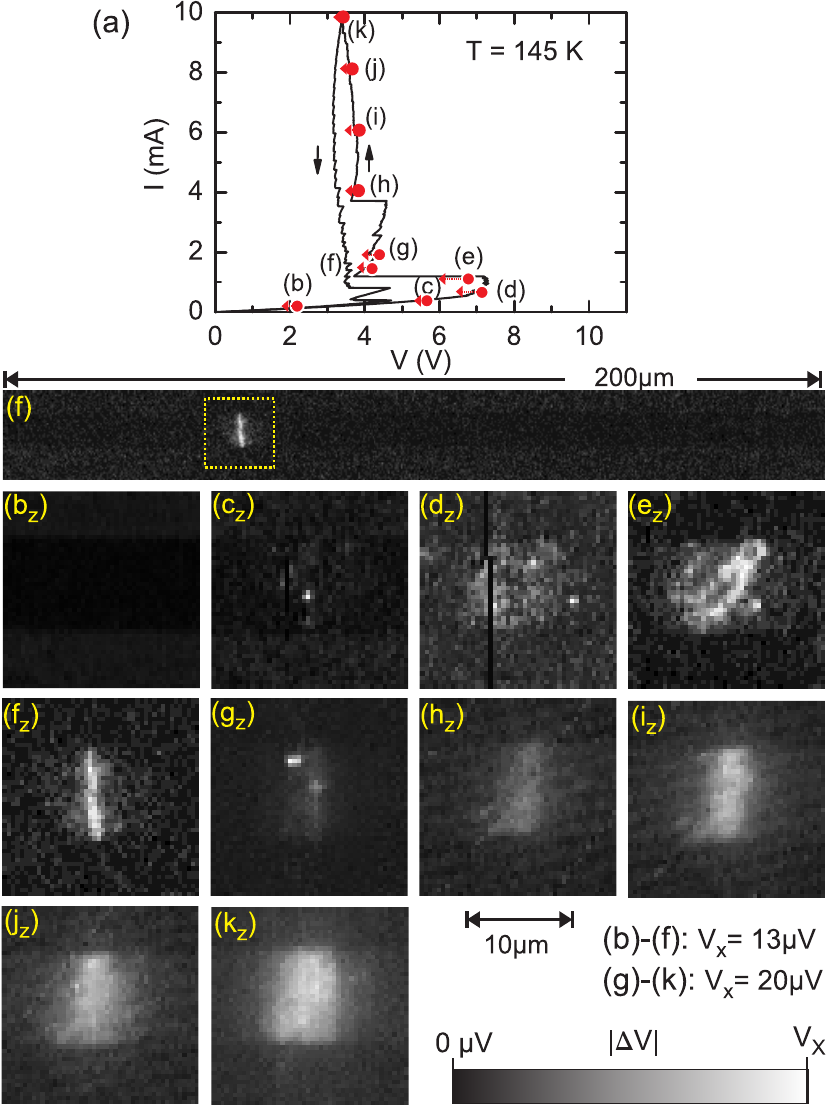}
\caption{\label{fig:sem145K}(Color online) 
LTSEM voltage-images acquired at different bias currents at a base temperature
of 145 K. 
(a) IV of figure 1  (c). 
The start points and the end points before and after imaging are indicated by
red arrows. 
(f) image of the entire device.
$(\textnormal{b}_z-\textnormal{k}_z)$  images with a reduced field of view
  (indicated by a yellow dashed rectangle in f).}
\end{figure}
%
The LTSEM images at a base temperature of 145\,K, are discussed below
[Fig. \ref{fig:sem145K}].
In order to relate the images to the current voltage characteristic, the IV of
the second sweep in Fig. \ref{fig:iv} is depicted.
We have discovered that the imaging process changes somewhat the resistance of the device. 
In Fig. \ref{fig:sem145K} (a) the bias points before and after imaging are
indicated by red arrows.   
In the images at bias points below the first big jump (b-e) a cluster of dots
appears in an approximately 10 $\mu\rm{m}$ wide section of the device indicated by a
dashed yellow rectangle in image (f). 
In the images at bias points above the first jump (f-k) a bright stripe appears
whose width increases with increasing currents.\\ 
%
\begin{figure}
\includegraphics[width=\linewidth]{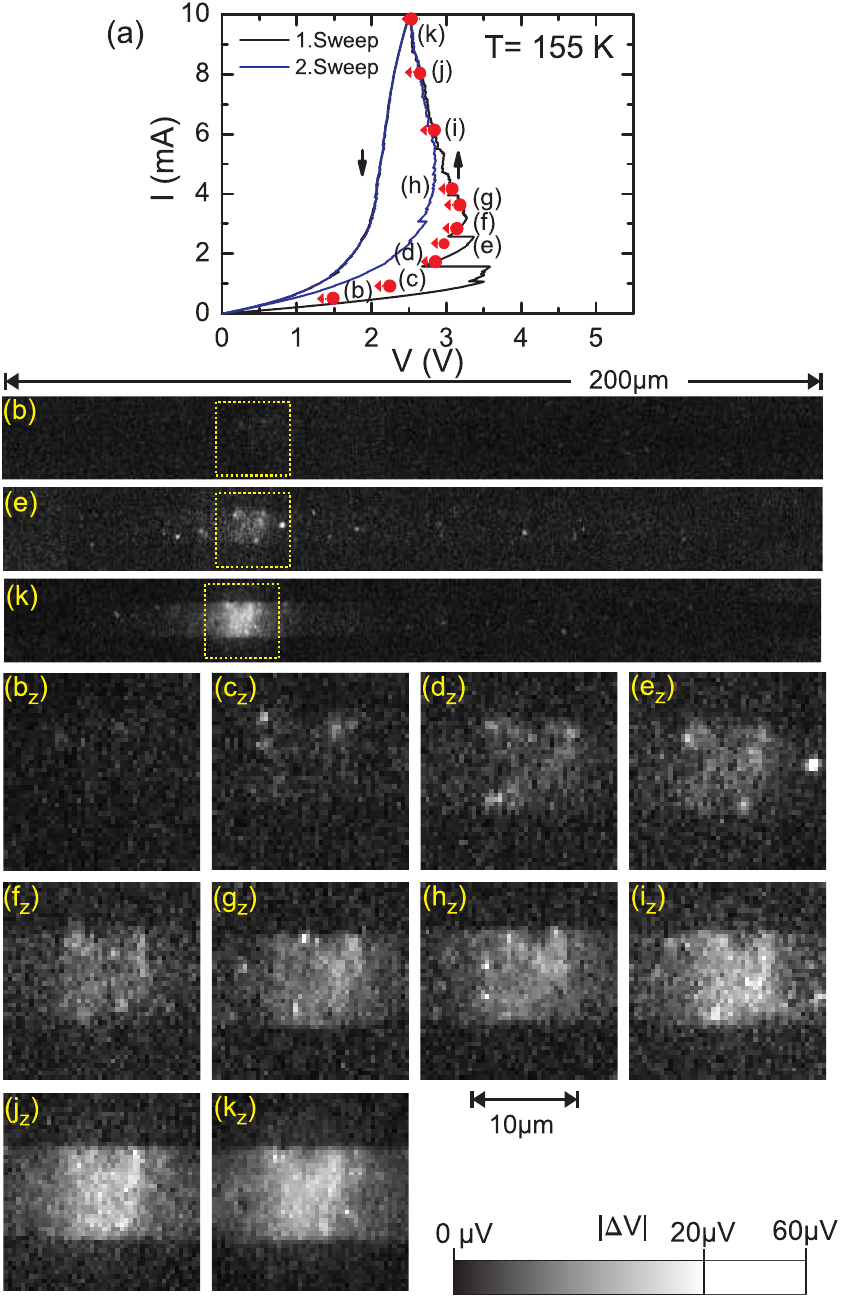}
\caption{\label{fig:sem155K}(Color online) 
LTSEM voltage images acquired at different bias currents at a base temperature
of 155 K.
(a) IVs of figure 1 (e-f).
The start points and the end points before and after imaging are
  indicated by red arrows. 
(b, e, and k) images of the entire device.
$(\textnormal{b}_z - \textnormal{k}_z)$ images with a reduced field of view
(indicated by a yellow dashed rectangle in b,e, and k).}
\end{figure}
%
A series of LTSEM images acquired
at a base temperature of $155\,$K are shown in Fig. \ref{fig:sem155K}. 
If the bias is increased, submicron sized dots of different brightness
appear. 
The majority of them is clustered in the same small section of
the device, where at $T_{b}=145\,$K the electrical breakdown occurs. 
For increasing currents the density of these dots increases and they merge.
When an imaging scan is repeated at the same bias current, the images look
more or less the same and additional bright spots appear at different positions.\\
In the supplement, the LTSEM response $\Delta V$ to the modulated electron beam
is estimated.
This shows that $\Delta V$ is proportional to the temperature derivative
of the conductivity close to the electron probe $\frac{dg}{dT}$.
Because $\frac{dg}{dT}$ in the insulating phase is several orders of magnitude smaller
than in the metallic phase, a large response $\Delta V$ is
predominately caused by the metallic domains.
The bright sub microns sized spots, which saturate the signal in the LTSEM
images, might be due to current redistribution.\\
In the LTSEM image series at T$_b$=145\,K (Fig. \ref{fig:sem145K}) a
metallic filament appears (electro-thermal domains) like reported in \vandi
devices \cite{Berglund1969, Duchene1971}.
In the T$_b$=155\,K series no filament appears, but the device
becomes gradually metallic in a small section of the device when the current
is increased.\\
\begin{figure*}[t]
\includegraphics{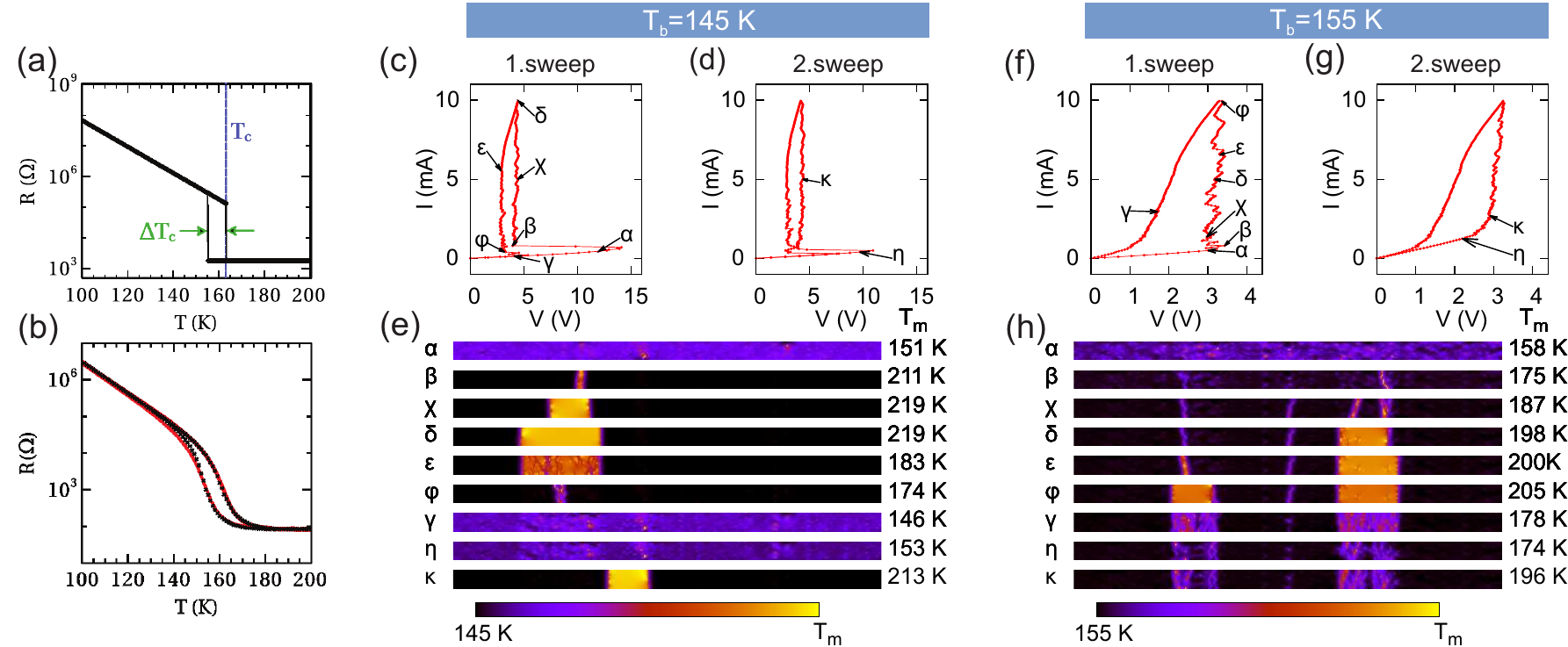}
\caption{\label{fig:simu}  
(Color Online)
Numerical model. 
(a) Assumed resistance vs. temperature dependence for an individual
domain. $T_c$ denotes the IMT temperature (heating curve). $\Delta T_c$ is the
temperature difference by which the cooling curve is shifted.
(b) Simulated resistance vs. temperature dependence. The crosses are
simulated points and the continuous line is the measured curve.
(c) Simulated IV of the pristine device (base temperature $T_b=145$\,K).
(d) Consecutive IV.
(e) Temperature distributions in the device at different bias points
(marked in (c) and (d) with Greek letters).
(f) Simulated IV of the pristine device (base temperature $T_b=155$\,K).
(g) Consecutive IV.
(h) Temperature distributions in the device at different bias points
(marked in (f) and (g) with Greek letters).
}
\end{figure*}
%
In order to investigate how the IMT transition of domains at the nanoscale are
influencing the electrical breakdown, we have developed a numerical model
(similar to the one used in \cite{Sharoni2008}),  in
which we represent domains with different IMT temperatures by a $20\times400$ resistor network (see supplement).  
For each domain the same hysteresis $\Delta T_c$ in the RT-dependence was assumed, but for
the IMT temperatures $T_c$ a Gaussian distribution was used
(Fig. \ref{fig:simu} a). 
The hysteresis as well as the median and the standard deviation of this
Gaussian distribution were obtained by optimizing the simulated RT-curve with
respect to the measured RT (Fig. \ref{fig:simu} b). 
Then these values ($\Delta T_c=8$\,K, $T_c=163$\,K  and RMSD\,$\approx3.16$)
were used to simulated the IVs as well as the temperature,
current and voltage distributions for the base temperatures $T_b=145\,$K
(Fig. \ref{fig:simu} c-e) and  $T_b=155\,$K (Fig. \ref{fig:simu} f-h). 
The simulated current and voltage distributions are shown in the supplement.\\
The hysteresis and the training effect in the IVs could be reproduced. 
The simulation at $T_b=145\,$K shows the formation electro-thermal domains
like in the LTSEM-images (Fig. \ref{fig:sem145K}).  
There are significant differences between the simulation and measurements at
$T_b=155\,$K: 
The simulation clearly shows the formation of electro-thermal domains
(filaments), while the IV progresses vertically (Fig. \ref{fig:simu} f-g). 
In the LTSLM images no filament appears and the IV progresses diagonally
(Fig. \ref{fig:sem155K}).\\ 
%
%
Conclusion: The hysteresis and the training effect in the IV are the results of
the RT hysteresis as well as the distribution of IMT temperatures
of nanoscaled domains in the polycrystalline thin film.
It is plausible that the RT hysteresis originates from the
structural bistability reported earlier \cite{Bao1998, Tanaka2001, Pfalzer2006}
and the spread in the transition temperatures is most likely caused by differences
in the strain in the polycrystalline film \cite{Grygiel2008, Schuler1997}.\\
At a base temperature of 145\,K the overall IV and the ETDs (filaments)
depicted in Fig. \ref{fig:sem145K} (f-k) were reproduced in the simulation,
clearly indicating an electro-thermal breakdown of the device.\\
On the contrary, at a base temperature of 155\,K the electrical breakdown,
with the small metallic domains clustered in a small section of the device and the diagonal
progression of the IV (Fig. \ref{fig:sem155K}), is very atypical for 
an electro-thermal breakdown. 
Generally, in a system, where the electro-thermal bistability is
caused by a large decrease in the device resistance with increasing
temperature the electro-thermal domain walls are parallel to the current direction, i.e. in the sample under investigation, electro-thermal domains have always
the shape of a filament.
After electro-thermal domains have nucleated, the IV progresses
vertically, while the hot electro-thermal domains increase in size
\cite{Volkov69, Gurevich87} (see also the supplement).\\ 
We provide three different explanations for the electrical breakdown via stable
intermediate states at a base temperature of 155\,K.
First, it is possible to stabilize an electro-thermal bistable 
device in intermediate states by using a load resistor, which forces the IV to
progress along a diagonal load line during the electro-thermal breakdown \cite{Fisher1975}.
The device under test has a relatively high contact resistance of
approximately $80\,\Omega$.
If one assumes that this contact resistance is nonohmic and has a tunneling
characteristic, the sawtooth like discontinuities and the
diagonal progression of the IV at $155\,$K can be explained.
Second, an intrinsic shunting of small domains above a certain threshold voltage
that could be caused by Landau-Zener tunneling \cite{Oka2003, Okamoto2008,
  Eckstein2010, Heidrich-Meisner2010} might have a similar effect. 
Third, in a theoretical model considering voltage induced switching two
different electrical breakdown mechanism were predicted - bolt like and
percolative switching \cite{Shekhawat2011}.  
According to this results one might interpret the electrical breakdown at a base
temperature of 145\,K as bolt like and at 155\,K as percolative switching. 
In this model thermal coupling within the thin film and to the substrate was
not considered. 
When we explicitly included voltage induced switching in our model, we were
never able to observe percolative switching, but we always observed the
formation of a filament. 
The only effect was to shift the thermal breakdown to lower voltages.
This is quite plausible, because the sample heating effects occur at lower
voltages, in this case.
From the simulated voltage distribution in the supplement a threshold
voltage for a dielectric breakdown below $20\,\rm{kV}/\rm{cm}$ can be
inferred. This value would be very small. Therefore, we consider the first explanation as the most likely scenario. 

\section{Supplement}
\subsection{Self-heating, electrothermal bistability and the formation of
  electrothermal domains (ETDs)}
Self-heating, thermal bistability and the formation of
electrothermal domains (ETDs), are discussed in the following section. 
The primary cause for these phenomena is a strong change in the resistivity
vs. temperature characteristic of the material used for the device under
investigation. 
These phenomena can be observed in large number
of different kind of devices: 
There is the hot-spot in a superconducting
microbridge \cite{Eichele83, Gurevich87, Doenitz07}, the pinch caused by thermal
breakdown in a negative temperature coefficient thermistor (NTC)
\cite{Spenke36, Spenke36b}, 
the domains in a two valley semi-conductor used
in a Gunn-diode \cite{Volkov69}, the hot spot in a large BSCCO-mesa used as a
THz-emitter \cite{Wang09, Wang10, Guenon10, Gross2012} or the filaments in a \vandi device
\cite{Berglund1969, Zhang1993} at the metal-insulator transition, to name only a few examples. 
In the following, only the basics of this topic necessary to understand the supplemented
paper are provided. 
A more elaborate discussion can be found in the review articles \cite{Volkov69, Gurevich87}. 
\subsubsection{Self-heating}
In a device with finite electrical resistance the Joule heating leads to a
rise in the device temperature. This temperature increase depends on the thermal
coupling of the device to the environment. 
If the device resistance is temperature dependent (as it is usually the case),
this self-heating effect causes a deviation in the current voltage
characteristic (IV) from the linear (Ohmic) progression.
In other words, there will be bending in the IV.
If the device resistance changes strongly with temperature, this bending can be
very pronounced as it is demonstrated in the following.\\
If the temperature dependence of the device resistance and the
thermal coupling are known, and a uniform temperature distribution is assumed
the IV can be parameterized using a method similar to the graphical method
described in \cite{Busch21}.\\
The excess temperature $\vartheta$ of the device is determined by Ohm's law and Newton's law
of cooling:
\begin{equation}
V=R(\vartheta)\,I;\quad V\,I=A(\vartheta),
\end{equation}
where $A(\vartheta)$ is the heat transfer coefficient.
The parameterization of the IV is 
\begin{equation}
V=\sqrt{R(\vartheta)\,A(\vartheta)}\quad \textnormal{and} \quad
I=\sqrt{\frac{A(\vartheta}{R(\vartheta)}}.
\label{eqn:parameterization}
\end{equation}
This method is now applied to the \vantri device discussed in the article.
On the assumption that the thermal conductance of the device is solely
determined by the sapphire substrate and ``edge cooling'' can be neglected
the heat transfer coefficient can be calculated using the thermal conductivity $\kappa$
of sapphire 
\begin{equation}
A(\vartheta)=\frac{L\,W}{H}\int_{T_b}^{T_b+\vartheta}\kappa(T)\,dT
\end{equation}
Here $L$ is the length, $W$ the width and $H$ the height of the device and
$T_{b}$ is the base temperature.\\
\begin{figure}
\includegraphics[width=0.9\linewidth]{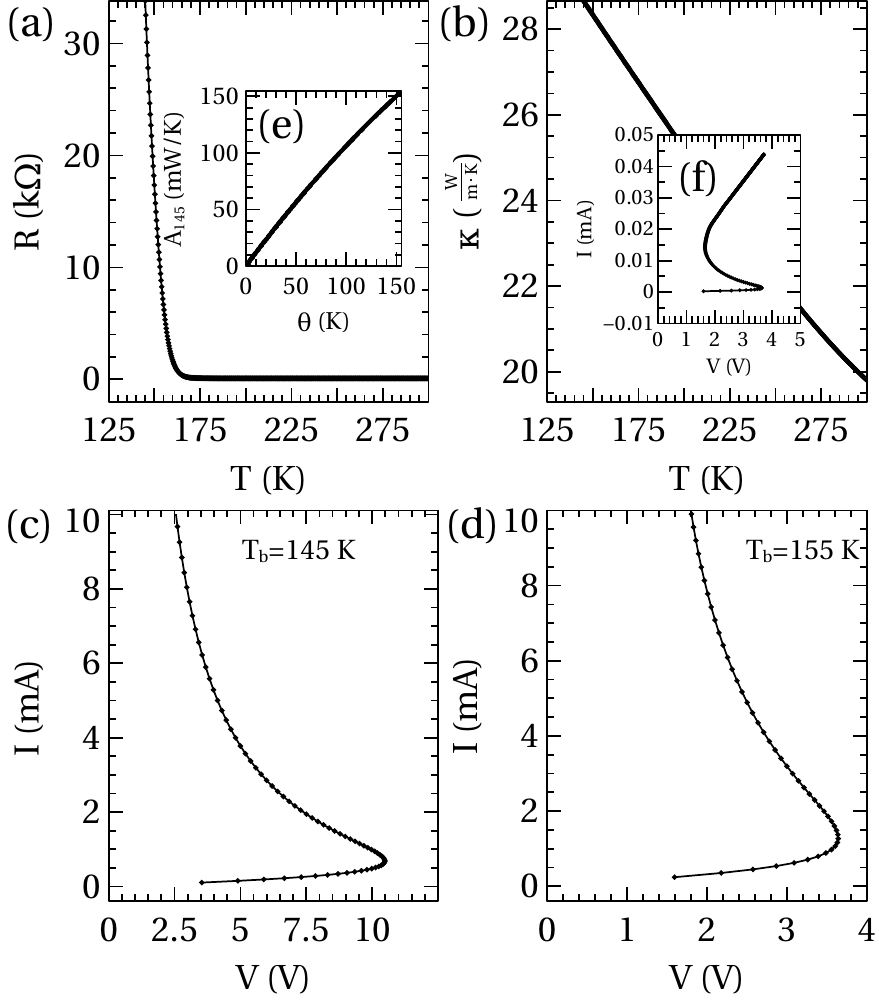}
\caption{\label{fig:simu_sub} 
Influence of self-heating on the current voltage characteristic of the
\vantri device: 
(a) Temperature dependence of the device resistance. 
(b) Thermal conductivity of sapphire vs. temperature. 
(c) and (d) calculated current voltage characteristics of the device, assuming
a homogenous temperature distribution. The base temperatures are $T_b=145\,K$ and $T_b=155\,K$, respectively. 
Inset (e) heat transfer coefficient via the sapphire substrate vs. excess temperature
  $\theta$ assuming a base temperature $T_b=145\,K$. 
Inset (f) IV of figure (d) at a larger scale. 
}
\end{figure}
Figure \ref{fig:simu_sub} c) and d) show the parameterized IV at the base
temperature $T_b=145\,\textnormal{K}$ and $T_b=155\,\textnormal{K}$,
respectively. 
$H=0.5\,\textnormal{mm}$, $W=200\,\mu\textnormal{m}$ and
$L=20\,\mu\textnormal{m}$ were used. 
%
%
For current values below $1\,\textnormal{mA}$ the calculated IVs of this
supplement are similar to the measured IVs presented in the main paper. 
The IV with a large current scale (inset [Fig. \ref{fig:simu_sub}(f)]) has a very
pronounced S-shaped characteristic. 
%
  \subsubsection{Electrothermal Bistability}
\begin{figure}
  \includegraphics[width=\linewidth]{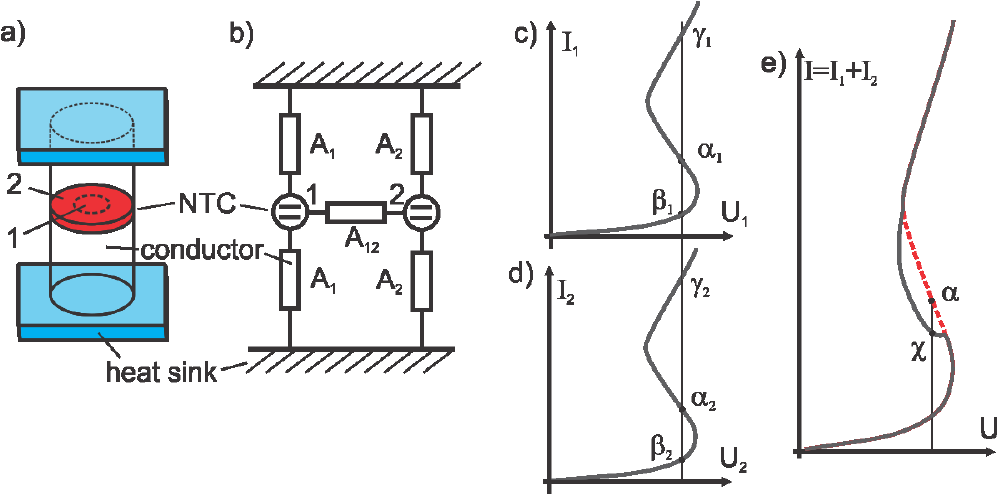}
  \caption{\label{fig:ETDs}
    Formation of electrothermal domains, according to \cite{Spenke36}:
    (a) NTC clamped between two metal pins.
    (b) Thermal equivalent circuit: A$_{1}$ and A$_{2}$ denote the thermal
    coupling to the heat sink and A$_{12}$ the thermal cross coupling, respectively.
    (c) and (d) IV curve corresponding to the area 1 and area 2 of the device.
    (e) IV of the whole device.}
\end{figure}

In the preceding section it was demonstrated that in a device, with a strong resistance decrease in a small temperature interval, self-heating leads to an S-shaped IV curve,
\cf [Fig. \ref{fig:simu_sub}(f)].
If a voltage source is used to bias such a device, then the system is
(electrothermally) bistable, i.e. there exists a bias interval in which the
system can rest in two different bias states:
The one state has a low bias current and a low device temperature (denoted in
\ref{fig:ETDs} (c) and (d) with $\beta_1$ and $\beta_2$), while the
other state has a high bias current and a high device temperature (denoted in
\ref{fig:ETDs} (c) and (d) with $\gamma_1$ and $\gamma_2$).
Note, that for a device with a strong increase in resistivity the
self-heating is causing an N-shaped IV and voltage and currents are changing their roles,
but this supplement is restricted on the case with an S-shaped IV.\\ 
\subsubsection{Formation of Electrothermal Domains}
Electrothermal domains can form in a device with weak thermal coupling
between different parts.  
%
%
According to \cite{Buettiker82} one of the first publications in the field of
ETDs was the work of E. Spenke on the thermal breakdown of negative
temperature coefficient thermistors (NTCs) \cite{Spenke36, Spenke36b}. 
He discussed a NTC disk fixed between two metal pins, which are in contact
with a thermal reservoir at a constant temperature, [Fig. \ref{fig:ETDs}(a)]. 
Note, that the resistance of the pins and the thermal reservoir is
considerably smaller than the resistance of the disk. 
In a ``Gedankenexperiment'',  E. Spenke divided the NTC-disk into two domains,
or in other words he considered the disk consisting of two NTCs in parallel, which are
thermally coupled via the metallic pins.
He inferred that the system (under current bias) can stabilize itself in the
bias interval of negative differential resistance, if one of those NTCs
switches to a bias point in the IV that corresponds to a high bias current and high
temperature (for instance bias point $\gamma_1$) and the other NTC switches to a bias point
with low current and low temperature (for instance bias point $\beta_2$). 
This is equivalent to the formation of two areas with different current
density and temperature, which are called electrothermal domains (ETDs).
Note, only if the ETD walls are parallel to the current direction, the ETDs
can be considered as parallel resistors and the argumentation of the
``Gedankenexperiment'' can be applied.\\
The formation of the electrothermal domains is causing a kink or a
discontinuity in the IV. 
If the S-shape in the IV of a device with uniform temperature distribution is
very pronounced than this discontinuity is very pronounced, too.
After the formation of the ETDs the IV progresses almost vertically indicating
that an increase of the current is increasing the size of the ``hot''
electrothermal domain with a high current density while the voltage over the
device stays constant. 

\subsection{Low temperature scanning electron microscopy (LTSEM)}

\begin{figure}
\includegraphics[width=0.75\linewidth]{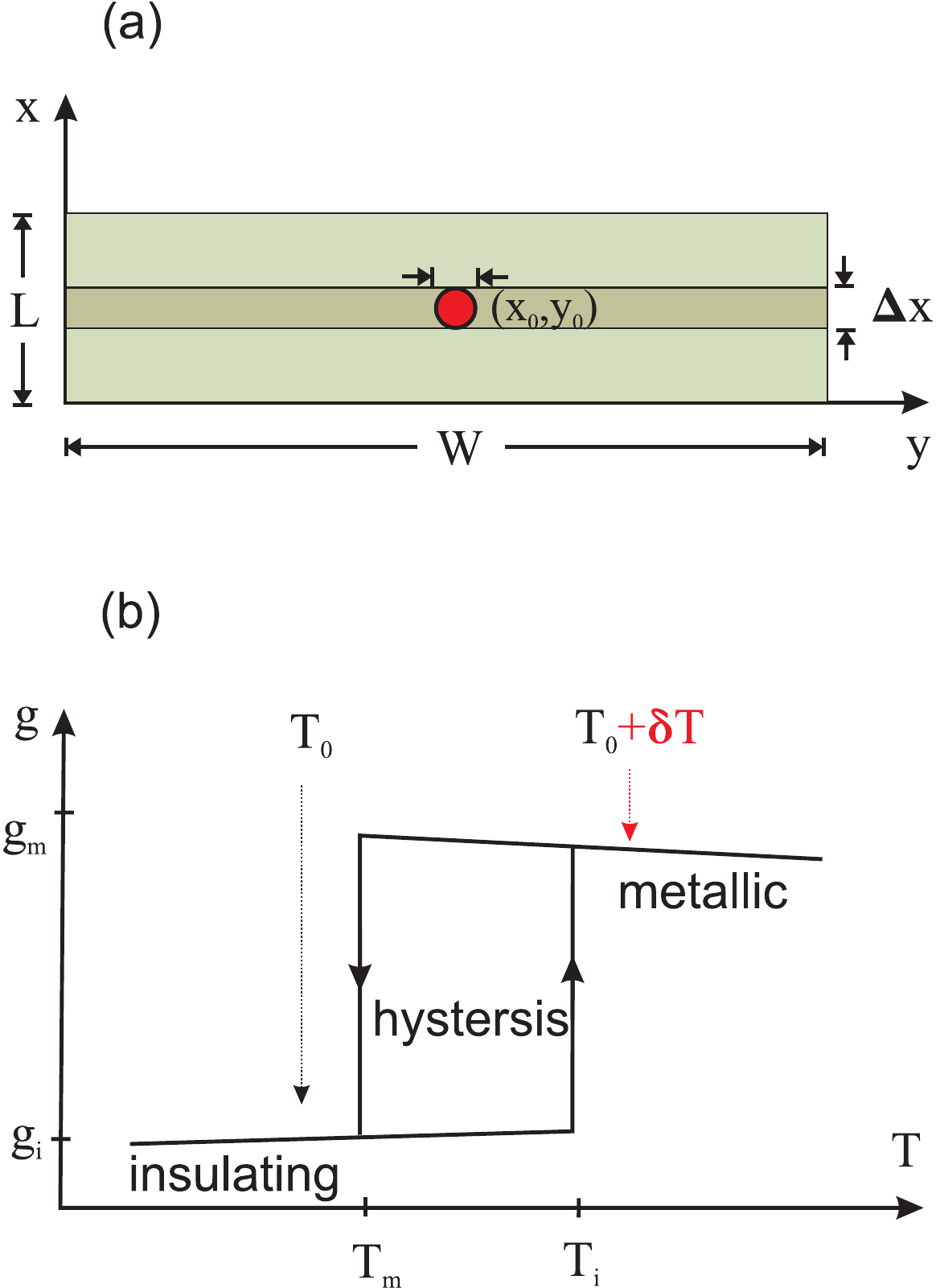}
\caption{\label{fig:DUI} 
a) Scheme of the device under investigation (top view), b) sketch of the specific conductivity vs. temperature. } 
\end{figure}

In the following we derive a relation between the LTSEM voltage signal $\Delta V(x_0,y_0)$ and the local conductance per area $g$ of the V$_2$O$_3$ sample.
In LTSEM a periodically blanked focused electron beam (using $f_b=13.3$\,kHz,
acceleration voltage $10$\,kV, beam current 100\,pA), is scanned across the sample surface in the ($x,y$)-plane. 
When the electron beam is positioned at coordinates ($x_0,y_0$), it causes local heating  and thus local changes in temperature-dependent parameters such as the local conductivity of the junction.
This can be detected by means of lock-in technique as a beam-induced change $\Delta V(x_0,y_0)$ of the voltage $V$ across the junction, which is biased at a constant current $I$.
The beam current also adds to the bias current density in the beam spot around ($x_0,y_0$). 
For measurements reported here, the beam current density is several orders of
magnitude smaller than the typical transport current densities. 
Therefore, this effect will be ignored here. 
\begin{widetext}
If the electron beam is on, the local temperature increases by $\delta T$  resulting in a temperature distribution $T(x-x_0,y-y_0)=T_0\,+\,\delta T(x-x_0,y-y_0)$, where $T_0$ is the local temperature of the undisturbed junction. The temperature profile created by the beam typically has Gaussian shape
\begin{equation}
  \delta T(x-x_0,y-y_0)=\Delta T \exp \left\{-\frac{(x-x_0)^2+(y-y_0)^2}{2\sigma ^2}\right\},
\label{Gauss}
\end{equation}
and $\Delta T \approx 0.1-1\,$K and $\sigma \approx 0.3-2$\,$\mu$m.
\end{widetext}
The geometry of the V$_2$O$_3$ device is a microbridge of length $L$ along
$x$, with a width $W$ along $y$ and a thickness $D$ along $z$ [Fig. \ref{fig:DUI} (a)].
Along $z$ the incident electron beam heats the bridge almost homogeneously and
we assume that the sample specific conductivity $g(x,y)$ does not depend on $z$. 
The bias current is applied in $x$ direction. 
Further, since $W \gg L$ we assume that the currents inside the bridge
strictly flow in $x$ direction, i. e. we neglect voltage drops in $y$
direction and current redistribution effects. 
With an electron beam illuminating a spot at the position $(x_0,y_0)$ the beam
induced voltage change is given by
\begin{equation}
  \label{eqn:deltaV}
  \Delta V(x_0,y_0)=I(R_{\rm{on}}-R_{\rm{off}}), 
\end{equation}
where $I$ is the total device current and $R_{\rm{on/off}}$ is the resistance when the
laser beam is on or off, respectively.
We divide the device in $N$ slabs oriented in the (y,z) plane with thickness
$\Delta x$. The $i$-th slap with the coordinate $x_i$ has the conductance
$G_i$ and the electron beam causes a small change in conductance $\delta
G_i$. The device resistance for the electron beam off state can be written as $R_{\rm{off}}=\sum_{i=1}^N
\frac{1}{G_i}$ and using a Taylor expansion 
\begin{equation}  
\label{eqn:r_taylor}
R_{\rm{on}}=\sum_{i=1}^N\frac{1}{G_i+\delta G_i}\approx \sum_{i=1}^N \frac{1}{G_i}-\sum_{i=1}^N\frac{\delta G_i}{G_i^2}
\end{equation}
From [\ref{eqn:deltaV}] and [\ref{eqn:r_taylor}] follows
\begin{equation}
\Delta V(x_0,y_0)=-I\,\sum_{i=1}^N\frac{\delta G_i}{G_i^2}
\end{equation}
We approximate the conductance of the $i$-th slab by averaging over the device
\begin{equation}
G_i\approx\frac{L}{\Delta x\,R_{\rm{off}}}.
\end{equation}
$\delta G_i$ can be written as 
\begin{equation}
\delta G_i=\frac{D}{\Delta x}\int_0^W\delta g(x_i,y,x_0,y_0)\,dy,
\end{equation}
where $\delta g(x,y,x_o,y_0)$ is the local change of conductivity cause by an
electron beam illuminating a spot at the position $(x_0, y_0)$.
\begin{widetext}
We obtain for the beam induced voltage change
\begin{equation}
\Delta V(x_0,y_0)= -\sum_{i=0}^N\int_0^W\frac{I\,D}{L^2}R_{\rm{off}}^2\delta
g(x_i,y,x_0,y_0)\,\Delta x\,dy =-\frac{I\,D}{L^2}R_{\rm{off}}^2 \int_A{\delta g(x,y,x_0,y_0)\,d(x,y)}
\end{equation}
\end{widetext}

The quantity to be evaluated further is $\int_A{\delta g(x,y)\,d(x,y)}$. 
In analogy with the assumed $R(T)$ dependence of an individual domain, see inset in
[Fig. 5] of the main paper, we consider $g(x,y)$ of the domain as a function of
$T$, which is characterized by three regions, see Fig. [\ref{fig:DUI} (b)].
At temperatures $T_0$ far away from the insulation-to-metal transition (IMT)
$g(x,y,T_0)$ is a unique function with $g_i$ for the insulating state and
$g_m$ for the metallic state, respectively. 
At the IMT the curve is hysteretic.\\  
The e-beam induced change in conductance of an individual domain is
qualitatively discussed in the following: 
$T_0$ is the temperature of the domain when the e-beam is off and $\delta T$
is the temperature increase of the domain when the e-beam is on. 
According to [Fig. \ref{fig:DUI}], if $T_0>T_i$ (where $T_i$ is the  insulator
to metal transition temperature) the change of the domain conductance is
proportional to the slope $\frac{dg_{m}}{dT}$. 
If $T_0<T_i$ and $T_0 + \delta T < T_i$, the change of the domain conductance
is proportional to the slope $\frac{dg_{i}}{dT}$ at $T_0$. 
If $T_0<T_i$ and $T_i<T_0+\delta T$, the illumination with the electron beam
causes an irreversible switching of the domain. 
This mechanism explains why the LTSEM imaging process is influencing the bias
over the device as discussed in the main paper. 
After the domain has switched the beam induced change of the conductance is
proportional to $\frac{dg_{m}}{dT}$.
Finally, if $T_0<T_m$ (where $T_m$ is the metal to insulator transition
temperature) and $T_i<T_0+\delta T$, the domain switches reversible between
the insulating and the metallic state and the beam induced change of the
domain conductance is proportional to $g_m-g_i\approx g_m$.
Due to the lock-in technique this response can be detected in LTSEM in
contrast to the irreversible switching.  
To summarize, there are three different response mechanism of an individual
domain: 
Metallic, insulating and the reversible switching, which is denoted with
the index $h$ (for hysteresis) in the following.\\
\begin{widetext}  
Because the device temperature locally varies and different domains have
different IMT temperatures, the device is divided in three areas
$A=A_m+A_i+A_h$. Note, that the areas are not necessarily connected. For
instance the device could consist of small metallic ``puddles'' embedded in an
insulating matrix.\\  
We rewrite  $\int_A{\delta g(x,y)\,d(x,y)}$ as   
  \begin{eqnarray}
  \int_A{\delta g(x,y)\,d(x,y)} = \int_{A_{m}}\left[{\frac{dg_m(x,y)}{dT}}|_{T_0}\,\delta T(x-x_0,y-y_0)\,\right]d(x,y)\label{g2}\\ 
  \nonumber +\int_{A_{i}}\left[{\frac{dg_i(x,y)}{dT}}|_{T_0}\,\delta T(x-x_0,y-y_0)\,\right]d(x,y)
   +\int_{A_{h}}{[g_m(x,y)-g_i(x,y)]\,d(x,y)}.
  \end{eqnarray}
\end{widetext}
The first term on the right hand side represents the regions of the sample
which remains in the metallic ($g_m$) state and the second term represents domains that remain in the insulating state ($g_i$). 
The third term represents the domains, which reversibly switch between the insulating and the metallic state upon illumination.
We may assume that $N$ domains in a radius $\approx \sigma$ switch between the
insulating and the metallic state. 
Let the area of the domain $n$ be $A_n$. Then the third term reduces to $\sum_n A_n \left[g_m(T_m)-g_i(T_i)\right]
\Theta(|\vec{r_0}-\vec{r_n}|/\sigma)$, where $\vec{r_0}$ is a vector to point
($x_0,y_0$) and $\vec{r_n}$ points to the center of the domain $n$.  
$\Theta(\xi)$ shall be 1 if $\xi<1$ and 0 otherwise. If $\sigma^2$ is on the order of or smaller than $A_n$ the domains which switch will be clearly distinguishable and $N$ = 0 or 1. For $\sigma^2 \gg A_n$ and not well separated switching domains one will see some blurred signal where the switching domains cannot be resolved clearly.\\ 
In the following we calculate and discuss the expected different responses of Eq. (\ref{g2}) for a device at 145 K at a bias 10\,mA and 3.5\,V which corresponds to image k in Fig. [3] of the main paper.  
For the insulating phase $g_i$ and $\frac{dg_i}{dT}$ can be established by means of the inset in Fig. [5] of the main paper. %
At $T=145\,$K we get $g_i\approx 26.6\frac{1}{\Omega\,\text{m}}$ and $\frac{dg_i}{dT} \approx 1.37\frac{1}{\Omega\,\text{K}\,\text{m}}$.
Because the device we discussed in the main paper has a high contact resistance, we used a four-terminal
measurement on a device with comparable \vantri thin film properties to determine $g_m\approx 1.43\cdot10^{5}\,\frac{1}{\Omega\,\text{m}}$ and $\frac{dg_{m}}{dT}\approx -310\frac{1}{\Omega\,\text{K}\,\text{m}}$.
%
%
\begin{widetext}
For the beam induced voltage change due to switching of domains we finally find
\begin{eqnarray}
  \Delta V_h(x_0,y_0) & \approx & -IR_{\rm{off}}^2\frac{D}{L^2}\sum_nA_n g_m(T_0) \Theta(|\vec{r_0}-\vec{r_n}|/\sigma)\\
   & \approx &  -0.05\,\rm{V},
\label{DeltaV1}
\end{eqnarray}
with $N=$1 and $A_n=(0.5\times0.5)\,\mu$m$^2$.
\end{widetext}
This expected value exceeds our measured signal by several orders of magnitude and is even higher for lower bias currents.  
According to the simulations of temperature distribution
[Fig. 6-7 of the main paper], the local temperature within an ETD ($T_{\rm{ETD}}\approx 210\,$K) is clearly above $T_i$. 
Furthermore, since the electron beam generates a local increase in temperature
of $\Delta T =0.25\,\rm{K}$ and the hysteresis has an amplitude of
$T_i-T_m=8$\,K\,$\gg \Delta T$, we rule out a periodic switching of domains.

%
%
\begin{widetext}
For the induced voltage change $\Delta V_{i,m}$ due to domains that remain in the insulating state or metallic state respectively, we find by using Eq. (\ref{Gauss}):
\begin{equation}
  \label{eqn:deltaV_approx}
 \Delta V_{i,m}(x_0,y_0)\approx-IR_{\rm{off}}^2\frac{D}{L^2}\Delta T \left\langle \frac{dg_{i,m}}{dT}(x_0,y_0) \right\rangle
\end{equation}
where $\left\langle \frac{dg_{i,m}}{dT}(x_0,y_0) \right\rangle$ is the
convolution of the local value of $\frac{dg_{i,m}}{dT}(x_0,y_0)$ with the
electron beam induced temperature profile. 
\end{widetext}
In order to obtain an upper estimate for the induced voltage change $\Delta
V$, we assume that an area $A_s=\frac{\pi}{4}(\Delta x)^2$ is homogeneously
heated up by the electron beam and that the entire area $A_s$ is either
metallic or insulating. 
\begin{widetext}
In this case equation \ref{eqn:deltaV_approx} reduces to 
\begin{equation}
  \Delta V_{i,m}(x_0,y_0)\approx -IR_{\rm{off}}^2\frac{D}{L^2}\Delta T A_s\frac{dg_{i,m}}{dT}(x_0,y_0)
\end{equation}
\end{widetext}
By considering the resolution of the LTSEM images in the main paper we can
estimate $\Delta x\approx 0.5\,\mu$m. 
We assume a beam induced temperature increase $\Delta T\approx 0.25\,$K. This
value is reasonable. For instance in \cite{Guerlich2010} $\Delta T=0.2-0.4\,$K
was obtained by considering a totally different LTSEM response mechanism. 
Finally, we find $\Delta V_{i}\approx -82\,$nV for the insulating phase (which may be neglected due to its smaller magnitude) and $\Delta V_{m}\approx 19\,\mu$V for the metallic phase.
From this we infer that mainly the metallic phase causes the response $\Delta
V$ in the LTSEM images.\\
Note, that the preceding estimate is an upper limit. 
The signal is reduced, if not all domains within the electron beam induced
temperature distribution are in the metallic state.
This explains the variation of brightness in the LTSEM images.
In addition, current redistribution effects are neglected in the preceding
model.
But these current redistribution effects can be considerable in a network of
fine conducting metallic filaments embedded in an insulating matrix.
This might be the reason for the very bright spots in the LTSEM images of
Fig. 4 in the main paper.
%
%
%
%
%
%

\begin{figure}
\includegraphics[width=\linewidth]{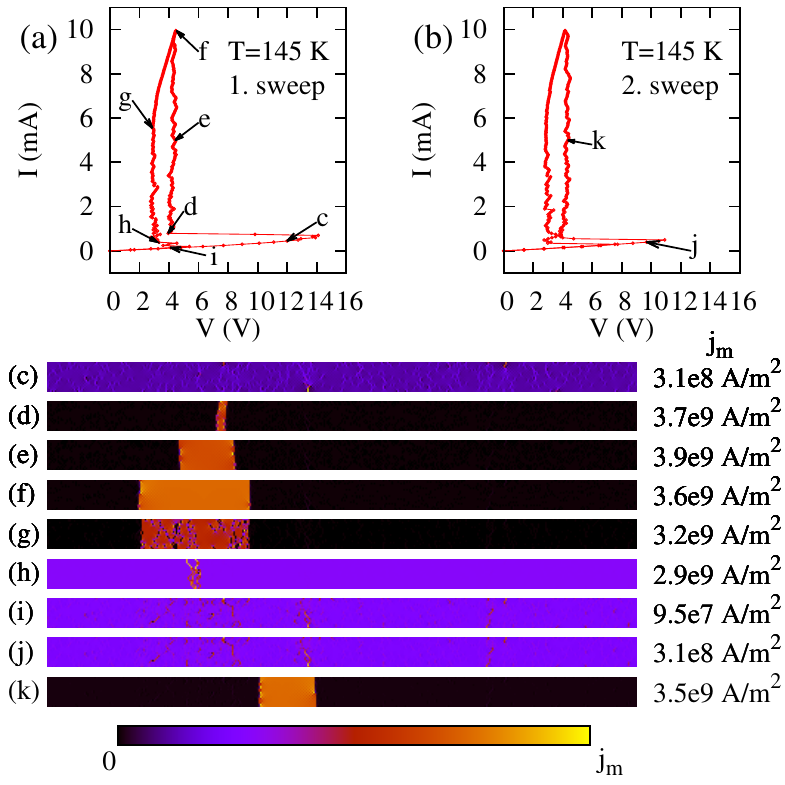}
\caption{\label{fig:simu145K_current} 
Simulation of the \vantri device with a base temperature of 145\,K. (a) IV of
the pristine device. (b) consecutive IV. (c-k) current distribution in the
device at different bias points (marked in Fig. a and b).}
\end{figure}

\begin{figure}
\includegraphics[width=\linewidth]{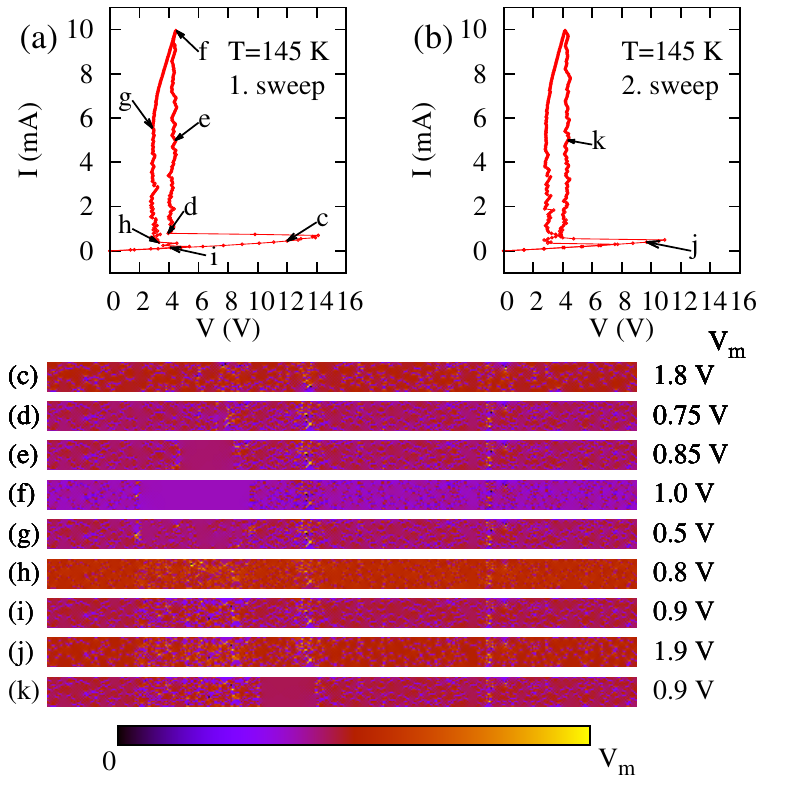}
\caption{\label{fig:simu145K_voltage} 
Simulation of the \vantri device with a base temperature of 145\,K. (a) IV of
the pristine device. (b) consecutive IV. (c-k) voltage distribution in the
device at different bias points (marked in Fig. a and b).}
\end{figure}


\begin{figure}
\includegraphics[width=\linewidth]{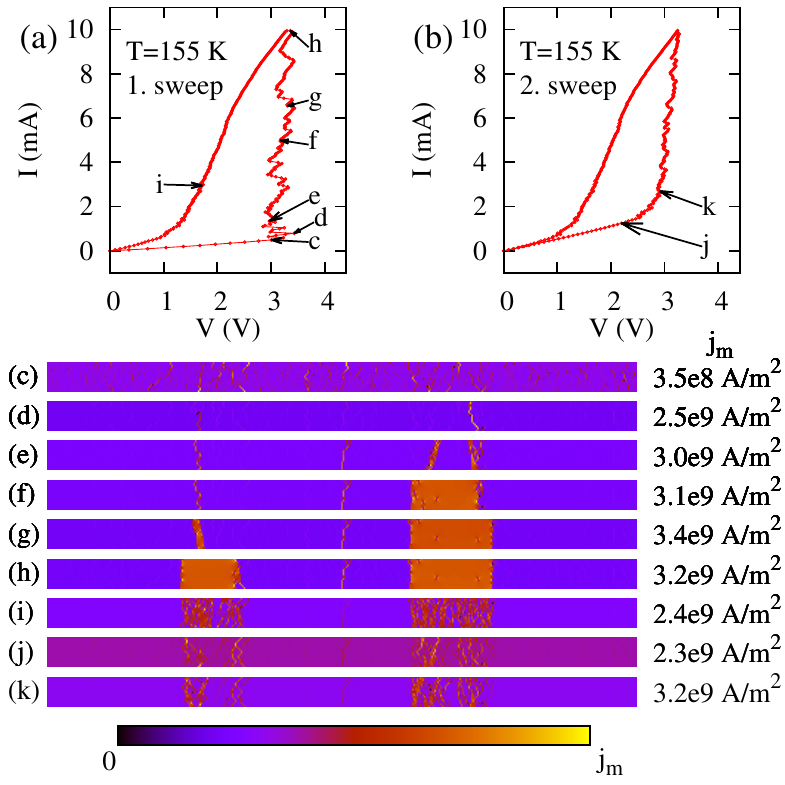}
\caption{\label{fig:simu155K_current}
Simulation of the \vantri device with a base temperature of 155\,K. (a) IV of
the pristine device. (b) consecutive IV. (c-k) current distribution in the
device at different bias points (marked in Fig. a and b)}
\end{figure}
\begin{figure}
\includegraphics[width=\linewidth]{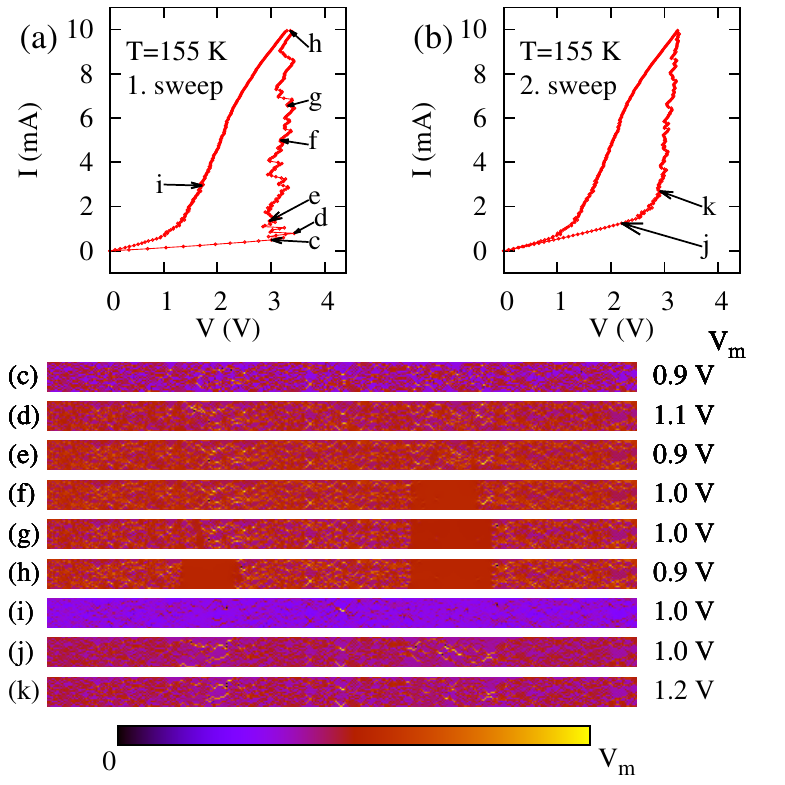}
\caption{\label{fig:simu155K_voltage}
Simulation of the \vantri device with a base temperature of 155\,K. (a) IV of
the pristine device. (b) consecutive IV. (c-k) voltage distribution in the
device at different bias points (marked in Fig. a and b)}
\end{figure}
%
\subsection{Computer Model and Additional Simulation Data}
\begin{figure}[t]
\includegraphics[width=\linewidth]{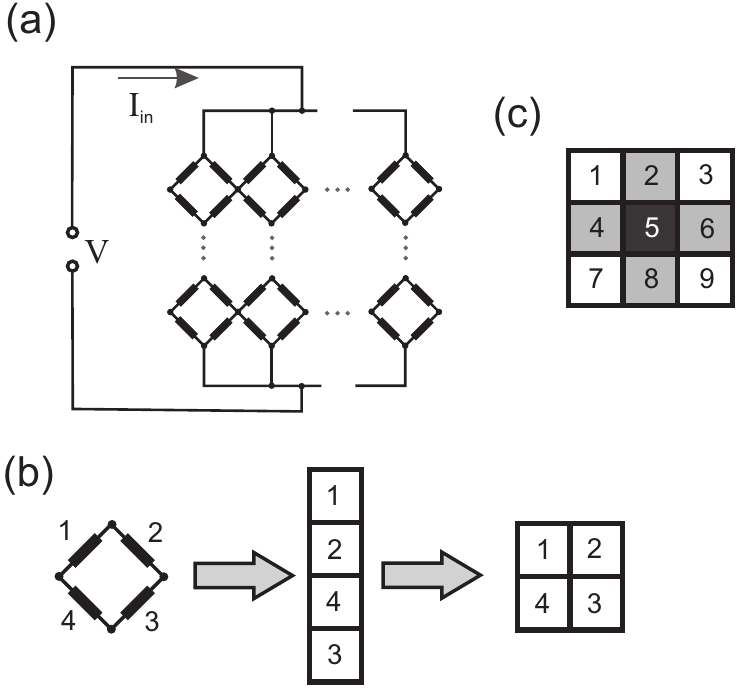}
\caption{\label{fig:diamond}
a) diamond shaped resistor network b) mapping of the resistor network to a
vector and  a 2D matrix c) thermal cross coupling to nearest neighbors.
}
\end{figure}

%
In order to simulate the Joule heating in the device, we have developed a
computer model, in which the domains are represented by a diamond shaped
resistor network [Fig. \ref{fig:diamond}(a)] (like in \cite{Shekhawat2011})
consisting of $M\times N$ diamonds.
Note, that the program we used is similar to the one used in
\cite{Sharoni2008}.
For the calculations, the electrical properties as well as the temperature of
these resistors are represented by $4\times M\times N$ dimensional vectors,
and for the visualization, we use two dimensional matrices [Fig. \ref{fig:diamond}(b)]. 
To simulate the IV, the program starts with an input current $I_{\rm{in}}=0$
and a homogenous temperature distribution $\vec{T}$ equals to the base
temperature $T_b$. 
For every resistor a temperature dependence of the resistance like in the
inset of Fig. 4 in the main paper with individual transition
temperatures is used.   
The conductance vector $\vec{G}$ is calculated using the temperature
distribution $\vec{T}$. 
We have developed an algorithm that creates a set of $4\times M\times N$
independent Kirchhoff equations corresponding to the resistor network
\begin{equation}
{\bf KM}(\vec{G})\vec{V}=\vec{I}_{\rm{in}}.
\end{equation}
We solve for the voltage distribution
\begin{equation}
\vec{V}={\bf KM}^{-1}\vec{I}_{\rm{in}},
\end{equation}
and calculate the current and power distributions
\begin{equation}
\vec{I}=\vec{G}\cdot\vec{V}, \quad \vec{P}=\vec{I}\cdot\vec{V}.
\end{equation}
With the power distribution the temperature distribution can be
recalculated.
For the thermal cross coupling (coupling parameter $A_c$), we consider only
coupling between nearest neighbors. 
For instance, for a network consisting of only 
9 resistors (depicted in Fig. \ref{fig:diamond}(c)) we get a thermal coupling
equation
\begin{equation}
   A_c(T_5-T_4)+A_c(T_5-T_2)+\cdots+A_s(T_5-T_b)=P_5,
\end{equation}
where $A_S$ is the thermal coupling parameter to the cold plate.
We have programmed an algorithm that creates the thermal coupling matrix ${\bf
  TCM}$.
The thermal coupling equations can be written as
\begin{equation}
 {\bf TCM}\,\vec{T}=\vec{P}+A_s\,T_b\,\mathds{1}.
\end{equation}
We solve for the temperature distribution
\begin{equation}
 \vec{T}=\left({\bf TCM}\right)^{-1}(\vec{P}+A_s\,T_b\,\mathds{1})
\end{equation}
We test for convergence by comparing the new temperature distribution with the
old one. 
If the solution is within the convergence criteria, we increment the input
current $I_{\rm in}$ and calculate the next bias point, otherwise we
recalculate the conductance vector $\vec{G}$ with the new temperature
distribution and iterate the program.\\
The RT curve is simulated in a similar manner. 
In this case the input current is fixed at a small value and the base
temperature is changed by a small increment.\\ 
The thermal cross coupling parameter
$A_c=5.0\cdot10^{-7}\frac{\text{W}}{\text{K}}$ was determined by considering
the heat flow in the \vantri film, only. 
The coupling parameter to the
cold plate $A_S=4.25\cdot10^{-7}\,\frac{\text{W}}{\text{K}}$ was obtained by optimizing
the simulation of the IV curves with respect to the measured IV.
The thermal conductance of a sapphire block with a ground area of $0.25\,
\mu\text{m}^2$ and a hight of $0.5\,$mm is approximately
$7.5\cdot10^{-8}\,\frac{\text{W}}{\text{K}}$. Here, it was assumed that
sapphire at a temperature $T=150\,$K has a thermal
conductivity of $150\,\frac{\text{W}}{\text{m}\cdot\text{K}}$ \cite{Touloukian1970}.  
If one takes into account that edge cooling effects were neglected, the
thermal coupling parameter $A_S$ is in reasonable agreement with this estimation.\\
In addition to the temperature distributions presented in the paper the voltage and
current distribution was simulated as well.
The current distributions depicted in Fig. (\ref{fig:simu145K_current}) and
  (\ref{fig:simu155K_current}) are very similar to the temperature
distributions presented in the main paper.\\ 
The voltage distributions are shown in Fig. (\ref{fig:simu145K_voltage}) and
  (\ref{fig:simu155K_voltage}).\\ 
%

\begin{acknowledgments}
This work was supported by AFOSR grant number FA9550-12-1-0381. 
For the computer simulation we used the linear algebra library 
Armadillo \cite{Sanderson2010}.
We gratefully acknowledge the pioneering work of R. P. Huebener in the field of
low temperature scanning electron microscopy. 
\end{acknowledgments}
\bibliography{SEM_V2O3}

\begin{thebibliography}{35}%
\makeatletter
\providecommand \@ifxundefined [1]{%
 \@ifx{#1\undefined}
}%
\providecommand \@ifnum [1]{%
 \ifnum #1\expandafter \@firstoftwo
 \else \expandafter \@secondoftwo
 \fi
}%
\providecommand \@ifx [1]{%
 \ifx #1\expandafter \@firstoftwo
 \else \expandafter \@secondoftwo
 \fi
}%
\providecommand \natexlab [1]{#1}%
\providecommand \enquote  [1]{``#1''}%
\providecommand \bibnamefont  [1]{#1}%
\providecommand \bibfnamefont [1]{#1}%
\providecommand \citenamefont [1]{#1}%
\providecommand \href@noop [0]{\@secondoftwo}%
\providecommand \href [0]{\begingroup \@sanitize@url \@href}%
\providecommand \@href[1]{\@@startlink{#1}\@@href}%
\providecommand \@@href[1]{\endgroup#1\@@endlink}%
\providecommand \@sanitize@url [0]{\catcode `\\12\catcode `\$12\catcode
  `\&12\catcode `\#12\catcode `\^12\catcode `\_12\catcode `\%12\relax}%
\providecommand \@@startlink[1]{}%
\providecommand \@@endlink[0]{}%
\providecommand \url  [0]{\begingroup\@sanitize@url \@url }%
\providecommand \@url [1]{\endgroup\@href {#1}{\urlprefix }}%
\providecommand \urlprefix  [0]{URL }%
\providecommand \Eprint [0]{\href }%
\providecommand \doibase [0]{http://dx.doi.org/}%
\providecommand \selectlanguage [0]{\@gobble}%
\providecommand \bibinfo  [0]{\@secondoftwo}%
\providecommand \bibfield  [0]{\@secondoftwo}%
\providecommand \translation [1]{[#1]}%
\providecommand \BibitemOpen [0]{}%
\providecommand \bibitemStop [0]{}%
\providecommand \bibitemNoStop [0]{.\EOS\space}%
\providecommand \EOS [0]{\spacefactor3000\relax}%
\providecommand \BibitemShut  [1]{\csname bibitem#1\endcsname}%
\let\auto@bib@innerbib\@empty
\bibitem [{\citenamefont {Yang}\ \emph {et~al.}(2011)\citenamefont {Yang},
  \citenamefont {Ko},\ and\ \citenamefont {Ramanathan}}]{Yang2011}%
  \BibitemOpen
  \bibfield  {author} {\bibinfo {author} {\bibfnamefont {Z.}~\bibnamefont
  {Yang}}, \bibinfo {author} {\bibfnamefont {C.}~\bibnamefont {Ko}}, \ and\
  \bibinfo {author} {\bibfnamefont {S.}~\bibnamefont {Ramanathan}},\ }\href
  {\doibase 10.1146/annurev-matsci-062910-100347} {\bibfield  {journal}
  {\bibinfo  {journal} {Annu. Rev. of Mater. Res.}\ }\textbf {\bibinfo {volume}
  {41}},\ \bibinfo {pages} {337} (\bibinfo {year} {2011})}\BibitemShut
  {NoStop}%
\bibitem [{\citenamefont {Oka}\ \emph {et~al.}(2003)\citenamefont {Oka},
  \citenamefont {Arita},\ and\ \citenamefont {Aoki}}]{Oka2003}%
  \BibitemOpen
  \bibfield  {author} {\bibinfo {author} {\bibfnamefont {T.}~\bibnamefont
  {Oka}}, \bibinfo {author} {\bibfnamefont {R.}~\bibnamefont {Arita}}, \ and\
  \bibinfo {author} {\bibfnamefont {H.}~\bibnamefont {Aoki}},\ }\href {\doibase
  10.1103/PhysRevLett.91.066406} {\bibfield  {journal} {\bibinfo  {journal}
  {Phys. Rev. Lett.}\ }\textbf {\bibinfo {volume} {91}},\ \bibinfo {pages}
  {066406} (\bibinfo {year} {2003})}\BibitemShut {NoStop}%
\bibitem [{\citenamefont {Okamoto}(2008)}]{Okamoto2008}%
  \BibitemOpen
  \bibfield  {author} {\bibinfo {author} {\bibfnamefont {S.}~\bibnamefont
  {Okamoto}},\ }\href {\doibase 10.1103/PhysRevLett.101.116807} {\bibfield
  {journal} {\bibinfo  {journal} {Phys. Rev. Lett.}\ }\textbf {\bibinfo
  {volume} {101}},\ \bibinfo {pages} {116807} (\bibinfo {year}
  {2008})}\BibitemShut {NoStop}%
\bibitem [{\citenamefont {Eckstein}\ \emph {et~al.}(2010)\citenamefont
  {Eckstein}, \citenamefont {Oka},\ and\ \citenamefont
  {Werner}}]{Eckstein2010}%
  \BibitemOpen
  \bibfield  {author} {\bibinfo {author} {\bibfnamefont {M.}~\bibnamefont
  {Eckstein}}, \bibinfo {author} {\bibfnamefont {T.}~\bibnamefont {Oka}}, \
  and\ \bibinfo {author} {\bibfnamefont {P.}~\bibnamefont {Werner}},\ }\href
  {\doibase 10.1103/PhysRevLett.105.146404} {\bibfield  {journal} {\bibinfo
  {journal} {Phys. Rev. Lett.}\ }\textbf {\bibinfo {volume} {105}},\ \bibinfo
  {pages} {146404} (\bibinfo {year} {2010})}\BibitemShut {NoStop}%
\bibitem [{\citenamefont {Heidrich-Meisner}\ \emph {et~al.}(2010)\citenamefont
  {Heidrich-Meisner}, \citenamefont {Gonz\'alez}, \citenamefont {Al-Hassanieh},
  \citenamefont {Feiguin}, \citenamefont {Rozenberg},\ and\ \citenamefont
  {Dagotto}}]{Heidrich-Meisner2010}%
  \BibitemOpen
  \bibfield  {author} {\bibinfo {author} {\bibfnamefont {F.}~\bibnamefont
  {Heidrich-Meisner}}, \bibinfo {author} {\bibfnamefont {I.}~\bibnamefont
  {Gonz\'alez}}, \bibinfo {author} {\bibfnamefont {K.~A.}\ \bibnamefont
  {Al-Hassanieh}}, \bibinfo {author} {\bibfnamefont {A.~E.}\ \bibnamefont
  {Feiguin}}, \bibinfo {author} {\bibfnamefont {M.~J.}\ \bibnamefont
  {Rozenberg}}, \ and\ \bibinfo {author} {\bibfnamefont {E.}~\bibnamefont
  {Dagotto}},\ }\href {\doibase 10.1103/PhysRevB.82.205110} {\bibfield
  {journal} {\bibinfo  {journal} {Phys. Rev. B}\ }\textbf {\bibinfo {volume}
  {82}},\ \bibinfo {pages} {205110} (\bibinfo {year} {2010})}\BibitemShut
  {NoStop}%
\bibitem [{\citenamefont {Liu}\ \emph {et~al.}(2012)\citenamefont {Liu},
  \citenamefont {Hwang}, \citenamefont {Tao}, \citenamefont {Strikwerda},
  \citenamefont {Fan}, \citenamefont {Keiser}, \citenamefont {Sternbach},
  \citenamefont {West}, \citenamefont {Kittiwatanakul}, \citenamefont {Lu},
  \citenamefont {Wolf}, \citenamefont {Omenetto}, \citenamefont {Zhang},
  \citenamefont {Nelson},\ and\ \citenamefont {Averitt}}]{Liu2012}%
  \BibitemOpen
  \bibfield  {author} {\bibinfo {author} {\bibfnamefont {M.}~\bibnamefont
  {Liu}}, \bibinfo {author} {\bibfnamefont {H.~Y.}\ \bibnamefont {Hwang}},
  \bibinfo {author} {\bibfnamefont {H.}~\bibnamefont {Tao}}, \bibinfo {author}
  {\bibfnamefont {A.~C.}\ \bibnamefont {Strikwerda}}, \bibinfo {author}
  {\bibfnamefont {K.}~\bibnamefont {Fan}}, \bibinfo {author} {\bibfnamefont
  {G.~R.}\ \bibnamefont {Keiser}}, \bibinfo {author} {\bibfnamefont {A.~J.}\
  \bibnamefont {Sternbach}}, \bibinfo {author} {\bibfnamefont {K.~G.}\
  \bibnamefont {West}}, \bibinfo {author} {\bibfnamefont {S.}~\bibnamefont
  {Kittiwatanakul}}, \bibinfo {author} {\bibfnamefont {J.}~\bibnamefont {Lu}},
  \bibinfo {author} {\bibfnamefont {S.~A.}\ \bibnamefont {Wolf}}, \bibinfo
  {author} {\bibfnamefont {F.~G.}\ \bibnamefont {Omenetto}}, \bibinfo {author}
  {\bibfnamefont {X.}~\bibnamefont {Zhang}}, \bibinfo {author} {\bibfnamefont
  {K.~A.}\ \bibnamefont {Nelson}}, \ and\ \bibinfo {author} {\bibfnamefont
  {R.~D.}\ \bibnamefont {Averitt}},\ }\href {\doibase 10.1038/nature11231}
  {\bibfield  {journal} {\bibinfo  {journal} {Nature}\ }\textbf {\bibinfo
  {volume} {487}},\ \bibinfo {pages} {345} (\bibinfo {year}
  {2012})}\BibitemShut {NoStop}%
\bibitem [{\citenamefont {Berglund}(1969)}]{Berglund1969}%
  \BibitemOpen
  \bibfield  {author} {\bibinfo {author} {\bibfnamefont {C.}~\bibnamefont
  {Berglund}},\ }\href {\doibase 10.1109/T-ED.1969.16773} {\bibfield  {journal}
  {\bibinfo  {journal} {IEEE Trans. Electron Devices}\ }\textbf {\bibinfo
  {volume} {16}},\ \bibinfo {pages} {432 } (\bibinfo {year}
  {1969})}\BibitemShut {NoStop}%
\bibitem [{\citenamefont {Duchene}\ \emph {et~al.}(1971)\citenamefont
  {Duchene}, \citenamefont {Terraillon}, \citenamefont {Pailly},\ and\
  \citenamefont {Adam}}]{Duchene1971}%
  \BibitemOpen
  \bibfield  {author} {\bibinfo {author} {\bibfnamefont {J.}~\bibnamefont
  {Duchene}}, \bibinfo {author} {\bibfnamefont {M.}~\bibnamefont {Terraillon}},
  \bibinfo {author} {\bibfnamefont {P.}~\bibnamefont {Pailly}}, \ and\ \bibinfo
  {author} {\bibfnamefont {G.}~\bibnamefont {Adam}},\ }\href {\doibase
  10.1063/1.1653835} {\bibfield  {journal} {\bibinfo  {journal} {Appl. Phys.
  Lett.}\ }\textbf {\bibinfo {volume} {19}},\ \bibinfo {pages} {115} (\bibinfo
  {year} {1971})}\BibitemShut {NoStop}%
\bibitem [{\citenamefont {Clem}\ and\ \citenamefont
  {Huebener}(1980)}]{Clem1980}%
  \BibitemOpen
  \bibfield  {author} {\bibinfo {author} {\bibfnamefont {J.~R.}\ \bibnamefont
  {Clem}}\ and\ \bibinfo {author} {\bibfnamefont {R.~P.}\ \bibnamefont
  {Huebener}},\ }\href {\doibase 10.1063/1.327939} {\bibfield  {journal}
  {\bibinfo  {journal} {J. Appl. Phys.}\ }\textbf {\bibinfo {volume} {51}},\
  \bibinfo {pages} {2764} (\bibinfo {year} {1980})}\BibitemShut {NoStop}%
\bibitem [{\citenamefont {Gross}\ and\ \citenamefont {Koelle}(1994)}]{Gross94}%
  \BibitemOpen
  \bibfield  {author} {\bibinfo {author} {\bibfnamefont {R.}~\bibnamefont
  {Gross}}\ and\ \bibinfo {author} {\bibfnamefont {D.}~\bibnamefont {Koelle}},\
  }\href@noop {} {\bibfield  {journal} {\bibinfo  {journal} {Rep. Prog. Phys.}\
  }\textbf {\bibinfo {volume} {57}},\ \bibinfo {pages} {651} (\bibinfo {year}
  {1994})}\BibitemShut {NoStop}%
\bibitem [{\citenamefont {Kim}\ \emph {et~al.}(2010)\citenamefont {Kim},
  \citenamefont {Ko}, \citenamefont {Frenzel}, \citenamefont {Ramanathan},\
  and\ \citenamefont {Hoffman}}]{Kim2010}%
  \BibitemOpen
  \bibfield  {author} {\bibinfo {author} {\bibfnamefont {J.}~\bibnamefont
  {Kim}}, \bibinfo {author} {\bibfnamefont {C.}~\bibnamefont {Ko}}, \bibinfo
  {author} {\bibfnamefont {A.}~\bibnamefont {Frenzel}}, \bibinfo {author}
  {\bibfnamefont {S.}~\bibnamefont {Ramanathan}}, \ and\ \bibinfo {author}
  {\bibfnamefont {J.~E.}\ \bibnamefont {Hoffman}},\ }\href {\doibase
  10.1063/1.3435466} {\bibfield  {journal} {\bibinfo  {journal} {Appl. Phys.
  Lett.}\ }\textbf {\bibinfo {volume} {96}},\ \bibinfo {eid} {213106} (\bibinfo
  {year} {2010})}\BibitemShut {NoStop}%
\bibitem [{\citenamefont {Grygiel}\ \emph {et~al.}(2008)\citenamefont
  {Grygiel}, \citenamefont {Pautrat}, \citenamefont {Prellier},\ and\
  \citenamefont {Mercey}}]{Grygiel2008}%
  \BibitemOpen
  \bibfield  {author} {\bibinfo {author} {\bibfnamefont {C.}~\bibnamefont
  {Grygiel}}, \bibinfo {author} {\bibfnamefont {A.}~\bibnamefont {Pautrat}},
  \bibinfo {author} {\bibfnamefont {W.}~\bibnamefont {Prellier}}, \ and\
  \bibinfo {author} {\bibfnamefont {B.}~\bibnamefont {Mercey}},\ }\href
  {http://stacks.iop.org/0295-5075/84/i=4/a=47003} {\bibfield  {journal}
  {\bibinfo  {journal} {EPL (Europhysics Letters)}\ }\textbf {\bibinfo {volume}
  {84}},\ \bibinfo {pages} {47003} (\bibinfo {year} {2008})}\BibitemShut
  {NoStop}%
\bibitem [{\citenamefont {Sharoni}\ \emph {et~al.}(2008)\citenamefont
  {Sharoni}, \citenamefont {Ram\'irez},\ and\ \citenamefont
  {Schuller}}]{Sharoni2008}%
  \BibitemOpen
  \bibfield  {author} {\bibinfo {author} {\bibfnamefont {A.}~\bibnamefont
  {Sharoni}}, \bibinfo {author} {\bibfnamefont {J.~G.}\ \bibnamefont
  {Ram\'irez}}, \ and\ \bibinfo {author} {\bibfnamefont {I.~K.}\ \bibnamefont
  {Schuller}},\ }\href {\doibase 10.1103/PhysRevLett.101.026404} {\bibfield
  {journal} {\bibinfo  {journal} {Phys. Rev. Lett.}\ }\textbf {\bibinfo
  {volume} {101}},\ \bibinfo {pages} {026404} (\bibinfo {year}
  {2008})}\BibitemShut {NoStop}%
\bibitem [{\citenamefont {Bao}\ \emph {et~al.}(1998)\citenamefont {Bao},
  \citenamefont {Broholm}, \citenamefont {Aeppli}, \citenamefont {Carter},
  \citenamefont {Dai}, \citenamefont {Rosenbaum}, \citenamefont {Honig},
  \citenamefont {Metcalf},\ and\ \citenamefont {Trevino}}]{Bao1998}%
  \BibitemOpen
  \bibfield  {author} {\bibinfo {author} {\bibfnamefont {W.}~\bibnamefont
  {Bao}}, \bibinfo {author} {\bibfnamefont {C.}~\bibnamefont {Broholm}},
  \bibinfo {author} {\bibfnamefont {G.}~\bibnamefont {Aeppli}}, \bibinfo
  {author} {\bibfnamefont {S.~A.}\ \bibnamefont {Carter}}, \bibinfo {author}
  {\bibfnamefont {P.}~\bibnamefont {Dai}}, \bibinfo {author} {\bibfnamefont
  {T.~F.}\ \bibnamefont {Rosenbaum}}, \bibinfo {author} {\bibfnamefont {J.~M.}\
  \bibnamefont {Honig}}, \bibinfo {author} {\bibfnamefont {P.}~\bibnamefont
  {Metcalf}}, \ and\ \bibinfo {author} {\bibfnamefont {S.~F.}\ \bibnamefont
  {Trevino}},\ }\href {\doibase 10.1103/PhysRevB.58.12727} {\bibfield
  {journal} {\bibinfo  {journal} {Phys. Rev. B}\ }\textbf {\bibinfo {volume}
  {58}},\ \bibinfo {pages} {12727} (\bibinfo {year} {1998})}\BibitemShut
  {NoStop}%
\bibitem [{\citenamefont {Tanaka}(2002)}]{Tanaka2001}%
  \BibitemOpen
  \bibfield  {author} {\bibinfo {author} {\bibfnamefont {A.}~\bibnamefont
  {Tanaka}},\ }\href {\doibase 10.1143/JPSJ.71.1091} {\bibfield  {journal}
  {\bibinfo  {journal} {J. Phys. Soc. Jpn.}\ }\textbf {\bibinfo {volume}
  {71}},\ \bibinfo {pages} {1091} (\bibinfo {year} {2002})}\BibitemShut
  {NoStop}%
\bibitem [{\citenamefont {Pfalzer}\ \emph {et~al.}(2006)\citenamefont
  {Pfalzer}, \citenamefont {Obermeier}, \citenamefont {Klemm}, \citenamefont
  {Horn},\ and\ \citenamefont {denBoer}}]{Pfalzer2006}%
  \BibitemOpen
  \bibfield  {author} {\bibinfo {author} {\bibfnamefont {P.}~\bibnamefont
  {Pfalzer}}, \bibinfo {author} {\bibfnamefont {G.}~\bibnamefont {Obermeier}},
  \bibinfo {author} {\bibfnamefont {M.}~\bibnamefont {Klemm}}, \bibinfo
  {author} {\bibfnamefont {S.}~\bibnamefont {Horn}}, \ and\ \bibinfo {author}
  {\bibfnamefont {M.~L.}\ \bibnamefont {denBoer}},\ }\href {\doibase
  10.1103/PhysRevB.73.144106} {\bibfield  {journal} {\bibinfo  {journal} {Phys.
  Rev. B}\ }\textbf {\bibinfo {volume} {73}},\ \bibinfo {pages} {144106}
  (\bibinfo {year} {2006})}\BibitemShut {NoStop}%
\bibitem [{\citenamefont {Schuler}\ \emph {et~al.}(1997)\citenamefont
  {Schuler}, \citenamefont {Klimm}, \citenamefont {Weissmann}, \citenamefont
  {Renner},\ and\ \citenamefont {Horn}}]{Schuler1997}%
  \BibitemOpen
  \bibfield  {author} {\bibinfo {author} {\bibfnamefont {H.}~\bibnamefont
  {Schuler}}, \bibinfo {author} {\bibfnamefont {S.}~\bibnamefont {Klimm}},
  \bibinfo {author} {\bibfnamefont {G.}~\bibnamefont {Weissmann}}, \bibinfo
  {author} {\bibfnamefont {C.}~\bibnamefont {Renner}}, \ and\ \bibinfo {author}
  {\bibfnamefont {S.}~\bibnamefont {Horn}},\ }\href {\doibase
  10.1016/S0040-6090(96)09399-6} {\bibfield  {journal} {\bibinfo  {journal}
  {Thin Solid Films}\ }\textbf {\bibinfo {volume} {299}},\ \bibinfo {pages}
  {119 } (\bibinfo {year} {1997})}\BibitemShut {NoStop}%
\bibitem [{\citenamefont {Volkov}\ and\ \citenamefont
  {Kogan}(1969)}]{Volkov69}%
  \BibitemOpen
  \bibfield  {author} {\bibinfo {author} {\bibfnamefont {A.}~\bibnamefont
  {Volkov}}\ and\ \bibinfo {author} {\bibfnamefont {M.}~\bibnamefont {Kogan}},\
  }\href@noop {} {\bibfield  {journal} {\bibinfo  {journal} {Sov. Phys. Usp.}\
  }\textbf {\bibinfo {volume} {11}},\ \bibinfo {pages} {881} (\bibinfo {year}
  {1969})}\BibitemShut {NoStop}%
\bibitem [{\citenamefont {Gurevich}\ and\ \citenamefont
  {Mints}(1987)}]{Gurevich87}%
  \BibitemOpen
  \bibfield  {author} {\bibinfo {author} {\bibfnamefont {A.~V.}\ \bibnamefont
  {Gurevich}}\ and\ \bibinfo {author} {\bibfnamefont {R.~G.}\ \bibnamefont
  {Mints}},\ }\href@noop {} {\bibfield  {journal} {\bibinfo  {journal} {Rev.
  Mod. Phys.}\ }\textbf {\bibinfo {volume} {59}},\ \bibinfo {pages} {941}
  (\bibinfo {year} {1987})}\BibitemShut {NoStop}%
\bibitem [{\citenamefont {Fisher}(1975)}]{Fisher1975}%
  \BibitemOpen
  \bibfield  {author} {\bibinfo {author} {\bibfnamefont {B.}~\bibnamefont
  {Fisher}},\ }\href {http://stacks.iop.org/0022-3719/8/i=13/a=016} {\bibfield
  {journal} {\bibinfo  {journal} {J. Phys. C}\ }\textbf {\bibinfo {volume}
  {8}},\ \bibinfo {pages} {2072} (\bibinfo {year} {1975})}\BibitemShut
  {NoStop}%
\bibitem [{\citenamefont {Shekhawat}\ \emph {et~al.}(2011)\citenamefont
  {Shekhawat}, \citenamefont {Papanikolaou}, \citenamefont {Zapperi},\ and\
  \citenamefont {Sethna}}]{Shekhawat2011}%
  \BibitemOpen
  \bibfield  {author} {\bibinfo {author} {\bibfnamefont {A.}~\bibnamefont
  {Shekhawat}}, \bibinfo {author} {\bibfnamefont {S.}~\bibnamefont
  {Papanikolaou}}, \bibinfo {author} {\bibfnamefont {S.}~\bibnamefont
  {Zapperi}}, \ and\ \bibinfo {author} {\bibfnamefont {J.~P.}\ \bibnamefont
  {Sethna}},\ }\href {\doibase 10.1103/PhysRevLett.107.276401} {\bibfield
  {journal} {\bibinfo  {journal} {Phys. Rev. Lett.}\ }\textbf {\bibinfo
  {volume} {107}},\ \bibinfo {pages} {276401} (\bibinfo {year}
  {2011})}\BibitemShut {NoStop}%
\bibitem [{\citenamefont {Eichele}\ \emph {et~al.}(1983)\citenamefont
  {Eichele}, \citenamefont {Freytag}, \citenamefont {Seifert}, \citenamefont
  {Huebener},\ and\ \citenamefont {Clem}}]{Eichele83}%
  \BibitemOpen
  \bibfield  {author} {\bibinfo {author} {\bibfnamefont {R.}~\bibnamefont
  {Eichele}}, \bibinfo {author} {\bibfnamefont {L.}~\bibnamefont {Freytag}},
  \bibinfo {author} {\bibfnamefont {H.}~\bibnamefont {Seifert}}, \bibinfo
  {author} {\bibfnamefont {R.~P.}\ \bibnamefont {Huebener}}, \ and\ \bibinfo
  {author} {\bibfnamefont {J.~R.}\ \bibnamefont {Clem}},\ }\href@noop {}
  {\bibfield  {journal} {\bibinfo  {journal} {J. Low Temp. Phys.}\ }\textbf
  {\bibinfo {volume} {52}},\ \bibinfo {pages} {449} (\bibinfo {year}
  {1983})}\BibitemShut {NoStop}%
\bibitem [{\citenamefont {Doenitz}\ \emph {et~al.}(2007)\citenamefont
  {Doenitz}, \citenamefont {Kleiner}, \citenamefont {Koelle}, \citenamefont
  {Scherer},\ and\ \citenamefont {Schuster}}]{Doenitz07}%
  \BibitemOpen
  \bibfield  {author} {\bibinfo {author} {\bibfnamefont {D.}~\bibnamefont
  {Doenitz}}, \bibinfo {author} {\bibfnamefont {R.}~\bibnamefont {Kleiner}},
  \bibinfo {author} {\bibfnamefont {D.}~\bibnamefont {Koelle}}, \bibinfo
  {author} {\bibfnamefont {T.}~\bibnamefont {Scherer}}, \ and\ \bibinfo
  {author} {\bibfnamefont {K.~F.}\ \bibnamefont {Schuster}},\ }\href {\doibase
  10.1063/1.2751109} {\bibfield  {journal} {\bibinfo  {journal} {Applied
  Physics Letters}\ }\textbf {\bibinfo {volume} {90}},\ \bibinfo {eid} {252512}
  (\bibinfo {year} {2007})}\BibitemShut {NoStop}%
\bibitem [{\citenamefont {Spenke}(1936{\natexlab{a}})}]{Spenke36}%
  \BibitemOpen
  \bibfield  {author} {\bibinfo {author} {\bibfnamefont {E.}~\bibnamefont
  {Spenke}},\ }\href@noop {} {\bibfield  {journal} {\bibinfo  {journal}
  {Electrical Engineering (Archiv für Elektrotechnik)}\ }\textbf {\bibinfo
  {volume} {30}},\ \bibinfo {pages} {728} (\bibinfo {year}
  {1936}{\natexlab{a}})}\BibitemShut {NoStop}%
\bibitem [{\citenamefont {Spenke}(1936{\natexlab{b}})}]{Spenke36b}%
  \BibitemOpen
  \bibfield  {author} {\bibinfo {author} {\bibfnamefont {E.}~\bibnamefont
  {Spenke}},\ }\href@noop {} {\bibfield  {journal} {\bibinfo  {journal}
  {Wissenschaftliche Veröffentlichungen aus den Siemens-Werken}\ }\textbf
  {\bibinfo {volume} {15}},\ \bibinfo {pages} {92} (\bibinfo {year}
  {1936}{\natexlab{b}})}\BibitemShut {NoStop}%
\bibitem [{\citenamefont {Wang}\ \emph {et~al.}(2009)\citenamefont {Wang},
  \citenamefont {Gu\'enon}, \citenamefont {Yuan}, \citenamefont {Iishi},
  \citenamefont {Arisawa}, \citenamefont {Hatano}, \citenamefont {Yamashita},
  \citenamefont {Koelle},\ and\ \citenamefont {Kleiner}}]{Wang09}%
  \BibitemOpen
  \bibfield  {author} {\bibinfo {author} {\bibfnamefont {H.~B.}\ \bibnamefont
  {Wang}}, \bibinfo {author} {\bibfnamefont {S.}~\bibnamefont {Gu\'enon}},
  \bibinfo {author} {\bibfnamefont {J.}~\bibnamefont {Yuan}}, \bibinfo {author}
  {\bibfnamefont {A.}~\bibnamefont {Iishi}}, \bibinfo {author} {\bibfnamefont
  {S.}~\bibnamefont {Arisawa}}, \bibinfo {author} {\bibfnamefont
  {T.}~\bibnamefont {Hatano}}, \bibinfo {author} {\bibfnamefont
  {T.}~\bibnamefont {Yamashita}}, \bibinfo {author} {\bibfnamefont
  {D.}~\bibnamefont {Koelle}}, \ and\ \bibinfo {author} {\bibfnamefont
  {R.}~\bibnamefont {Kleiner}},\ }\href {\doibase
  10.1103/PhysRevLett.102.017006} {\bibfield  {journal} {\bibinfo  {journal}
  {Phys. Rev. Lett.}\ }\textbf {\bibinfo {volume} {102}},\ \bibinfo {pages}
  {017006} (\bibinfo {year} {2009})}\BibitemShut {NoStop}%
\bibitem [{\citenamefont {Wang}\ \emph {et~al.}(2010)\citenamefont {Wang},
  \citenamefont {Gu\'enon}, \citenamefont {Gross}, \citenamefont {Yuan},
  \citenamefont {Jiang}, \citenamefont {Zhong}, \citenamefont {Gr\"unzweig},
  \citenamefont {Iishi}, \citenamefont {Wu}, \citenamefont {Hatano},
  \citenamefont {Koelle},\ and\ \citenamefont {Kleiner}}]{Wang10}%
  \BibitemOpen
  \bibfield  {author} {\bibinfo {author} {\bibfnamefont {H.~B.}\ \bibnamefont
  {Wang}}, \bibinfo {author} {\bibfnamefont {S.}~\bibnamefont {Gu\'enon}},
  \bibinfo {author} {\bibfnamefont {B.}~\bibnamefont {Gross}}, \bibinfo
  {author} {\bibfnamefont {J.}~\bibnamefont {Yuan}}, \bibinfo {author}
  {\bibfnamefont {Z.~G.}\ \bibnamefont {Jiang}}, \bibinfo {author}
  {\bibfnamefont {Y.~Y.}\ \bibnamefont {Zhong}}, \bibinfo {author}
  {\bibfnamefont {M.}~\bibnamefont {Gr\"unzweig}}, \bibinfo {author}
  {\bibfnamefont {A.}~\bibnamefont {Iishi}}, \bibinfo {author} {\bibfnamefont
  {P.~H.}\ \bibnamefont {Wu}}, \bibinfo {author} {\bibfnamefont
  {T.}~\bibnamefont {Hatano}}, \bibinfo {author} {\bibfnamefont
  {D.}~\bibnamefont {Koelle}}, \ and\ \bibinfo {author} {\bibfnamefont
  {R.}~\bibnamefont {Kleiner}},\ }\href {\doibase
  10.1103/PhysRevLett.105.057002} {\bibfield  {journal} {\bibinfo  {journal}
  {Phys. Rev. Lett.}\ }\textbf {\bibinfo {volume} {105}},\ \bibinfo {pages}
  {057002} (\bibinfo {year} {2010})}\BibitemShut {NoStop}%
\bibitem [{\citenamefont {Gu\'enon}\ \emph {et~al.}(2010)\citenamefont
  {Gu\'enon}, \citenamefont {Gr\"unzweig}, \citenamefont {Gross}, \citenamefont
  {Yuan}, \citenamefont {Jiang}, \citenamefont {Zhong}, \citenamefont {Li},
  \citenamefont {Iishi}, \citenamefont {Wu}, \citenamefont {Hatano},
  \citenamefont {Mints}, \citenamefont {Goldobin}, \citenamefont {Koelle},
  \citenamefont {Wang},\ and\ \citenamefont {Kleiner}}]{Guenon10}%
  \BibitemOpen
  \bibfield  {author} {\bibinfo {author} {\bibfnamefont {S.}~\bibnamefont
  {Gu\'enon}}, \bibinfo {author} {\bibfnamefont {M.}~\bibnamefont
  {Gr\"unzweig}}, \bibinfo {author} {\bibfnamefont {B.}~\bibnamefont {Gross}},
  \bibinfo {author} {\bibfnamefont {J.}~\bibnamefont {Yuan}}, \bibinfo {author}
  {\bibfnamefont {Z.~G.}\ \bibnamefont {Jiang}}, \bibinfo {author}
  {\bibfnamefont {Y.~Y.}\ \bibnamefont {Zhong}}, \bibinfo {author}
  {\bibfnamefont {M.~Y.}\ \bibnamefont {Li}}, \bibinfo {author} {\bibfnamefont
  {A.}~\bibnamefont {Iishi}}, \bibinfo {author} {\bibfnamefont {P.~H.}\
  \bibnamefont {Wu}}, \bibinfo {author} {\bibfnamefont {T.}~\bibnamefont
  {Hatano}}, \bibinfo {author} {\bibfnamefont {R.~G.}\ \bibnamefont {Mints}},
  \bibinfo {author} {\bibfnamefont {E.}~\bibnamefont {Goldobin}}, \bibinfo
  {author} {\bibfnamefont {D.}~\bibnamefont {Koelle}}, \bibinfo {author}
  {\bibfnamefont {H.~B.}\ \bibnamefont {Wang}}, \ and\ \bibinfo {author}
  {\bibfnamefont {R.}~\bibnamefont {Kleiner}},\ }\href {\doibase
  10.1103/PhysRevB.82.214506} {\bibfield  {journal} {\bibinfo  {journal} {Phys.
  Rev. B}\ }\textbf {\bibinfo {volume} {82}},\ \bibinfo {pages} {214506}
  (\bibinfo {year} {2010})}\BibitemShut {NoStop}%
\bibitem [{\citenamefont {{Gross}}\ \emph {et~al.}(2012)\citenamefont
  {{Gross}}, \citenamefont {{Guenon}}, \citenamefont {{Yuan}}, \citenamefont
  {{Li}}, \citenamefont {{Li}}, \citenamefont {{Iishi}}, \citenamefont
  {{Mints}}, \citenamefont {{Hatano}}, \citenamefont {{Wu}}, \citenamefont
  {{Koelle}}, \citenamefont {{Wang}},\ and\ \citenamefont
  {{Kleiner}}}]{Gross2012}%
  \BibitemOpen
  \bibfield  {author} {\bibinfo {author} {\bibfnamefont {B.}~\bibnamefont
  {{Gross}}}, \bibinfo {author} {\bibfnamefont {S.}~\bibnamefont {{Guenon}}},
  \bibinfo {author} {\bibfnamefont {J.}~\bibnamefont {{Yuan}}}, \bibinfo
  {author} {\bibfnamefont {M.~Y.}\ \bibnamefont {{Li}}}, \bibinfo {author}
  {\bibfnamefont {J.}~\bibnamefont {{Li}}}, \bibinfo {author} {\bibfnamefont
  {A.}~\bibnamefont {{Iishi}}}, \bibinfo {author} {\bibfnamefont {R.~G.}\
  \bibnamefont {{Mints}}}, \bibinfo {author} {\bibfnamefont {T.}~\bibnamefont
  {{Hatano}}}, \bibinfo {author} {\bibfnamefont {P.~H.}\ \bibnamefont {{Wu}}},
  \bibinfo {author} {\bibfnamefont {D.}~\bibnamefont {{Koelle}}}, \bibinfo
  {author} {\bibfnamefont {H.~B.}\ \bibnamefont {{Wang}}}, \ and\ \bibinfo
  {author} {\bibfnamefont {R.}~\bibnamefont {{Kleiner}}},\ }\href@noop {}
  {\bibfield  {journal} {\bibinfo  {journal} {ArXiv e-prints}\ } (\bibinfo
  {year} {2012})},\ \Eprint {http://arxiv.org/abs/1206.6275} {arXiv:1206.6275
  [cond-mat.supr-con]} \BibitemShut {NoStop}%
\bibitem [{\citenamefont {Zhang}\ and\ \citenamefont
  {Eklund}(1993)}]{Zhang1993}%
  \BibitemOpen
  \bibfield  {author} {\bibinfo {author} {\bibfnamefont {J.-G.}\ \bibnamefont
  {Zhang}}\ and\ \bibinfo {author} {\bibfnamefont {P.}~\bibnamefont {Eklund}},\
  }\href@noop {} {\bibfield  {journal} {\bibinfo  {journal} {J. Mater. Res.}\ }
  (\bibinfo {year} {1993})}\BibitemShut {NoStop}%
\bibitem [{\citenamefont {Busch}(1921)}]{Busch21}%
  \BibitemOpen
  \bibfield  {author} {\bibinfo {author} {\bibfnamefont {H.}~\bibnamefont
  {Busch}},\ }\href@noop {} {\bibfield  {journal} {\bibinfo  {journal} {Annalen
  der Physik}\ }\textbf {\bibinfo {volume} {64}},\ \bibinfo {pages} {401}
  (\bibinfo {year} {1921})}\BibitemShut {NoStop}%
\bibitem [{\citenamefont {Büttiger}\ and\ \citenamefont
  {Landauer}(1982)}]{Buettiker82}%
  \BibitemOpen
  \bibfield  {author} {\bibinfo {author} {\bibfnamefont {M.}~\bibnamefont
  {Büttiger}}\ and\ \bibinfo {author} {\bibfnamefont {R.}~\bibnamefont
  {Landauer}},\ }in\ \href@noop {} {\emph {\bibinfo {booktitle} {Nonlinear
  Phenomena at Phase Transitions and Instabilities}}},\ \bibinfo {editor}
  {edited by\ \bibinfo {editor} {\bibfnamefont {T.}~\bibnamefont {Riste}}}\
  (\bibinfo  {publisher} {Plenum, New York/London},\ \bibinfo {year} {1982})\
  p.\ \bibinfo {pages} {111}\BibitemShut {NoStop}%
\bibitem [{\citenamefont {G\"urlich}\ \emph {et~al.}(2010)\citenamefont
  {G\"urlich}, \citenamefont {Scharinger}, \citenamefont {Weides},
  \citenamefont {Kohlstedt}, \citenamefont {Mints}, \citenamefont {Goldobin},
  \citenamefont {Koelle},\ and\ \citenamefont {Kleiner}}]{Guerlich2010}%
  \BibitemOpen
  \bibfield  {author} {\bibinfo {author} {\bibfnamefont {C.}~\bibnamefont
  {G\"urlich}}, \bibinfo {author} {\bibfnamefont {S.}~\bibnamefont
  {Scharinger}}, \bibinfo {author} {\bibfnamefont {M.}~\bibnamefont {Weides}},
  \bibinfo {author} {\bibfnamefont {H.}~\bibnamefont {Kohlstedt}}, \bibinfo
  {author} {\bibfnamefont {R.~G.}\ \bibnamefont {Mints}}, \bibinfo {author}
  {\bibfnamefont {E.}~\bibnamefont {Goldobin}}, \bibinfo {author}
  {\bibfnamefont {D.}~\bibnamefont {Koelle}}, \ and\ \bibinfo {author}
  {\bibfnamefont {R.}~\bibnamefont {Kleiner}},\ }\href {\doibase
  10.1103/PhysRevB.81.094502} {\bibfield  {journal} {\bibinfo  {journal} {Phys.
  Rev. B}\ }\textbf {\bibinfo {volume} {81}},\ \bibinfo {pages} {094502}
  (\bibinfo {year} {2010})}\BibitemShut {NoStop}%
\bibitem [{\citenamefont {Touloukian}\ and\ \citenamefont
  {Buyco}(1970)}]{Touloukian1970}%
  \BibitemOpen
  \bibfield  {author} {\bibinfo {author} {\bibfnamefont {Y.~S.}\ \bibnamefont
  {Touloukian}}\ and\ \bibinfo {author} {\bibfnamefont {E.~H.}\ \bibnamefont
  {Buyco}},\ }\href@noop {} {\emph {\bibinfo {title} {Thermal Conductivity}}},\
  Vol.\ \bibinfo {volume} {1 and 2}\ (\bibinfo  {publisher} {Plenum Press, New
  York},\ \bibinfo {year} {1970})\BibitemShut {NoStop}%
\bibitem [{\citenamefont {Sanderson}(2010)}]{Sanderson2010}%
  \BibitemOpen
  \bibfield  {author} {\bibinfo {author} {\bibfnamefont {C.}~\bibnamefont
  {Sanderson}},\ }\href@noop {} {\emph {\bibinfo {title} {{Armadillo: An Open
  Source C++ Linear Algebra Library for Fast Prototyping and Computationally
  Intensive Experiments}}}},\ \bibinfo {type} {Tech. Rep.}\ (\bibinfo
  {institution} {NICTA},\ \bibinfo {address} {Australia},\ \bibinfo {year}
  {2010})\BibitemShut {NoStop}%
\end{thebibliography}%

\end{document}